\documentclass[journal]{IEEEtran}
\usepackage{amsmath,bm}
\usepackage{enumerate}
\usepackage{amssymb}
\usepackage[mathscr]{eucal}
\usepackage{relsize}
\usepackage{hyperref}
\usepackage{array}
\usepackage{graphicx}
\usepackage{balance}
\usepackage{array}
\usepackage{url}
\usepackage{caption,subcaption,gensymb}
\usepackage{tabularx}
\usepackage{graphicx}
\usepackage{cite}
\usepackage{graphics} % for pdf, bitmapped graphics files
\usepackage{algorithm}
\usepackage{algpseudocode}
\usepackage{xcolor}

\newcommand\norm[1]{\lVert#1\rVert}
\newcommand\red[1]{\textcolor{red}{#1}}

\begin{document}
\title{Coverage Path Planning with Budget Constraints for Multiple Unmanned Ground Vehicles}

\author{Vu Phi Tran, Asanka Perera, Matthew A. Garratt,~\IEEEmembership{Senior Member,~IEEE}, Kathryn Kasmarik,~\IEEEmembership{Senior Member,~IEEE}, Sreenatha Anavatti % <-this % stops a space
% \thanks{Vu Phi Tran, Asanka Perera, Matthew A. Garratt, Kathryn Kasmarik, and Sreenatha Anavatti are with the School of Engineering and Information Technology, University of New South Wales, Canberra, Australia. %Alex S. Leong, and Mohammad Zamani are with Land Division, Defence Science and Technology Group, Melbourne, Australia. 
% E-mail: \{phi.tran, asanka.perera, m.garratt, k.kasmarik, and s.anavatti\}@adfa.edu.au. This research is funded by the Australian Defence Science and Technology (DST) Group.}% <-this % stops a space
}

\maketitle

\begin{abstract}
This paper proposes an innovative approach to coverage path planning and obstacle avoidance for multiple Unmanned Ground Vehicles (UGVs) in a changing environment, taking into account constraints on the time, path length, number of UGVs and obstacles. Our approach leverages deformable virtual leader-follower formations to enable UGVs to adapt their formation based on both planned and real-time sensor data. A hierarchical block algorithm is employed to identify areas in the environment where UGV formations can spread out to meet time and budget constraints. Additionally, we introduce a novel control scheme that allows each UGV to generate a local steering force to dodge any static and mobile obstacles based on the closest safe angle. Results from simulations and real UGV experiments demonstrate that our approach achieves a higher coverage percentage than rule-based and reactive swarming approaches without planning. Our approach offers a promising solution for efficient coverage path planning and obstacle avoidance in complex environments with multiple UGVs.

\end{abstract}
%The closed-loop control stability is verified through the Lyapunov theory.
\IEEEpeerreviewmaketitle

\section{Introduction}
Intelligent transportation is gaining popularity with an increase in the number of practical applications \cite{huang20213}.  This is particularly true for autonomous systems, such as unmanned ground vehicles (UGVs).  UGVs have different applications, including coverage of a given area. In particular, when a number of UGVs/agents are employed to cover a given area, the control systems need to be intelligent to achieve the mission while overcoming the obstacles, both static as well as dynamic. The coverage path planning problem is a task wherein a UGV or UGVs, possessing a complete geometric description of the area of interest, generates an efficient coverage path to visit every point in a given area while avoiding all possible obstacles \cite{galceran2013}. Various technological developments and advancements in sensor technology, navigational, communication, and computational systems have facilitated the rapid growth in the use of coverage path planning (CPP) methods to assist UGVs in performing many specific applications, ranging from humanitarian missions such as surveillance, search and rescue tasks,  to military operations such as surveillance \cite{wai2019}, environmental monitoring \cite{pham2017}, and civilian applications such as area cleaning, seeding or harvesting \cite{wang2022}, and mapping and model reconstruction \cite{liu2019}.

% In the last decade, several approaches have been discussed in the literature for coverage by a single vehicle \cite{niroui2019deep}. However, a single agent may not be able to meet an operational time limit or may be limited by real-world factors such as battery capacity or sensor payload restrictions \cite{almadhoun2019}. Compared with single-vehicle CPP, the larger footprint of a group of multiple vehicles should mean they can solve a coverage task more rapidly. However, novel algorithms are required to exploit the capacity of a multi-vehicle team by determining the route to be followed by each vehicle, when a group of autonomous vehicles can spread out, and when they must take on a narrow formation. 

In recent years, the literature has discussed several approaches for coverage by a single vehicle \cite{niroui2019deep}. However, real-world factors such as battery capacity or sensor payload restrictions \cite{almadhoun2019} may limit the ability of a single agent to meet an operational time limit. Compared to single-vehicle CPP, a group of multiple vehicles may solve a coverage task more rapidly due to its larger footprint. Yet, to exploit the capacity of a multi-vehicle team, novel algorithms are required to determine the route for each vehicle when they can spread out and take on a narrow formation.

%However, utilizing a single robot to cover a large structure or a wide area can take a long time, might require the robot to recharge mid way, or it might be difficult for a robot to carry sensors to acquire all the information it needs

Motivated by the aforementioned observations, the contributions of the present paper are:
\begin{itemize}
	\item A novel problem definition for non-backtracking coverage path planning with  budget constraints assuming the use of a multi-UGV team in cluttered and uncertain environments.
  
    \item A novel algorithm is presented for solving the problem, which utilizes a hierarchical block approach to decompose a given map into appropriate cell sizes. This allows us to exploit flexible multi-UGV formations to meet multiple budget constraints.
    
    \item A distributed virtual leader-follower formation control strategy including automatic role assignment in the formation and obstacle avoidance.

    \item A comprehensive comparative study in both simulated and real-world settings to confirm the viability of our approach. Our approach outperforms existing methods in terms of maximum coverage percentage, time to achieve coverage and computational complexity.
    
\end{itemize}

The virtual leader-follower control approach taken in this paper is based on the virtual spring system \cite{colombo2021forced}. The virtual leader-follower approach offers several advantages over the traditional leader-follower and swarm-based methods in coverage path planning. Firstly, it eliminates the need for a physical leader, which can be costly and risky to implement. Secondly, it allows for the efficient coordination of multiple followers without the risk of collisions or formation breakage, which is a common issue with swarm-based approaches. Thirdly, a virtual leader can easily adapt to changes in the environment, dynamically adjust the path plan, and provide more accurate and reliable instructions to the followers. This is in contrast to the real leader-follower system, which may not be able to respond quickly enough to changes in the environment \cite{tran2021hybrid,tran2021multi}. Finally, the virtual leader-follower approach offers more flexibility and scalability, enabling the coordination of a large number of followers without requiring additional resources.

The remainder of this paper is organised as follows. Section II discusses related work from the literature. Section III states our coverage path planning problem definition and describes our approach to solving this problem. Our approach has components for path planning and prediction of how long it will take to follow a path in formation. Section IV presents a series of experiments with each of these components, first in simulations then on real UGVs in an outdoor setting. We offer conclusions and directions for future work in Section V.

\section{Literature Review}

Various techniques for area coverage exist, including Voronoi-based \cite{palacios2016}, graph-based \cite{palacios2019equitable}, next best view \cite{manjanna2018heterogeneous}, frontier-based \cite{tran2023dynamic}, and spanning tree coverage methods \cite{dong2020}. While some techniques provide incomplete coverage, others like spanning tree coverage guarantee complete coverage by generating a non-overlapping path along the spanning tree. In this study, we employ the spanning tree method for multi-unmanned ground vehicle (UGV) settings where several UGVs are used to cover an area.

Developing Multi-Cell Path Planning (MCPP) strategies for real-world scenarios is a challenging task that requires considering additional factors and requirements. Previous studies on MCPP have mainly focused on obstacle-free spaces \cite{shao2021bipartite,chen2021clustering} or single cover types \cite{gao2018}, which may not apply to many real-world scenarios. Only a block data structured method \cite{jan2019} has been developed so far to obtain a comprehensive and safe area coverage path for a team of vehicles, using the concepts of a contour map, connected graph, and spanning tree. However, this method may not scale well to large-scale multi-vehicle systems due to the tiny cells generated around obstacle boundaries. Additionally, no existing methods focus on dynamic environments with non-stationary obstacles, which are more common in practice. Therefore, detecting and avoiding moving obstacles correctly and efficiently is crucial.

Most MCPP algorithms rely on centralized control, which can lead to communication overhead and coordination difficulties as the number of agents increases \cite{zhou2022}. To address this, a distributed approach can be used to improve coordination and communication among vehicles, generate more efficient coverage paths, and better adapt to changes in the environment. Additionally, a distributed approach provides greater robustness to failures or disturbances (e.g., the loss of a vehicle, changes in the environment, or communication failures) by allowing the system to reconfigure dynamically \cite{tran2020}.

Furthermore, most MCPP algorithms, such as multi-robot forest coverage \cite{gorbenko2015multi} and spiral spanning tree coverage \cite{gao2019stc}, assume a desire for 100\% area coverage without recharging or refilling robots \cite{ma2022cciba}. However, in real-world applications, many physical constraints need to be considered, such as limited exploration time, travelled path length, restricted access to some areas, and different types of coverage required. Therefore, the state-of-the-art MCPP strategies compute plans that lead to imperfect coverage but are considerably more energy/time-efficient \cite{ellefsen2017}. Additionally, complete coverage may not even be feasible when areas to be covered have impeded or hidden components, further highlighting the need for partial coverage \cite{ellefsen2016}.

Several solutions have applied multi-objective evolutionary techniques in path planning to balance coverage and energy/time \cite{ellefsen2017,zhou2021}. However, such multi-objective approaches cannot provide a satisfactory solution when multiple objectives are considered, as in many real-world scenarios. To the best of our knowledge, optimizing the grid cell size, a key parameter in the MCPP problem with constraints, while meeting all specific goals and limitations, has not been extensively examined. Additionally, for large-scale workspaces, multi-objective evolutionary optimization techniques are often time-consuming due to offline training. Therefore, the presented methods are not appropriate for a dynamic setting, where real-time decisions must be made \cite{wen2021}. A simpler alternative with very low computational complexity must be taken into account.

Moreover, controlling a flexible leader-follower formation to track a pre-defined path is a relatively unexplored approach compared to allocating a specific area to each vehicle. However, this approach has significant benefits which we wish to exploit. By staying close together in a re-configurable formation, the robots can communicate and use visual relative positioning, which can be crucial in scenarios where GPS signals are degraded or denied. The leader-follower approach also allows the vehicles to adapt to changes in the environment or mission requirements, which is not possible with fixed area allocation. This approach also enables the formation to be more robust to failures or disturbances, as the leader can quickly adjust the formation to account for the loss of a vehicle, making the system more resilient \cite{tran2022frontier}.

Overall, the MCPP problem is a complex optimization problem that requires consideration of multiple factors and constraints, including area coverage, energy/time efficiency, communication overhead, robustness, obstacles, and uncertainty. While there have been many advancements in this field, there is still much work to be done to develop practical and effective MCPP strategies that can be applied to a wide range of real-world scenarios. In the next section, we formulate the problem addressed in this paper and present our solution. 

\section{Problem Definition and Algorithm}
In this section, we begin in Part A by describing the MCPP problem under the physical limits we address in this paper. We provide our algorithmic solution in Part B, followed by a complexity analysis in Part C.

\subsection{Problem Formulation}
The simple unicycle model captures the trade-off between linear and angular velocity control for small unmanned ground vehicles (UGVs) \cite{aghaeeyan2015} and forms the basis of the UGV simulation used in this paper. We denote $q$ as the UGV state comprising its position and heading, whilst  $u$ is the velocity command vector:
\begin{equation}
q=[x~y~\theta]^T \in \mathbb{R}^3, u = [v~\omega]^T \in \mathbb{R}^2.
\end{equation}
We will refer to two components of the state $q$: $q_p \triangleq [x~y]^T \in \mathbb{R}^2$, the position, and $q_\theta \triangleq \theta \in \mathbb{R}$ the yaw angle. The coordinates $x$ and $y$ capture the position of the center of gravity of the vehicle while $\theta$ is the angle of the wheel with respect to the $x$-axis of the reference coordinate frame. The control variable $v$ is the forward velocity of the vehicle (along its body fixed frame $x$-axis) while $\omega$ is the rate of change of the yaw angle.

The following equations capture the unicycle model with the known initial conditions $x_0$, $y_0$ and $\theta_0$:
\begin{equation}
\begin{split}
\dot{x} &= v~cos(\theta), x(0) = x_0, \\
\dot{y} &= v~cos(\theta), y(0) = y_0, \\
\dot{\theta} &= \omega, \theta(0) = \theta_0. \\
\end{split}
\end{equation}

We assume the environment's shape and obstacles are known. Consider an environment $\mathcal{C}$ to be explored by a formation of $n_q$ vehicles, $\mathcal{Q} = \{q_1,..., q_{n_q}\}$. We choose a tessellated environment of grid cells of width $C_W$ and height $C_H$ such that $\mathcal{C} = \{c_{ij} | i = 1,\dots,\mathcal{C}_W, j = 1,\dots,\mathcal{C}_H\}$. Given a discretised map $\mathcal{C}$ with a set of known obstacles $\mathcal{O}$, a leader-follower formation with initial configuration $Q^0 \in \mathcal{C}_{free}$ and $ \mathcal{C}_{free} \triangleq \mathcal{C} \setminus \mathcal{O}$, a time budget of $T_{max}$ steps, (and/or a UGV path length budget of $L_{max}$) and a set of observed cells at  time $k$, $\sigma^k := \sigma^{k-1} \cup \zeta^k,$ where $\zeta^k = \cup_{i=1}^{n_q}(\zeta_i^k)$ is the set of new cells covered by all UGVs' total observation area at the time $k$. Calculate a plan $\mathcal{P}^T := Q^0,..., Q^k,..., Q^T, \forall T \leq T_{max}$ that does not cross any obstacles $(Q^k \in \mathcal{C}_{free}, \forall k)$, and maximizes the coverage percentage $CP$ with respect to cell size $CS$. In other words, the MCCP problem with constraints requires us to maximise:
\begin{equation}\label{eq:problem}
CP=\frac{\sigma^T}{C_{free} + OB}  \times 100\%
\end{equation}
subject to:
\begin{equation}\label{eq:constraints}
\begin{split}
& \sum_{k=0}^{T} \norm{Q_p^k - Q_p^{k+1}}_2 \leq L_{max},\\
& \norm{Q_p^k - Q_p^{k+1}}_2 \leq v_{max}, \forall~k, and\\
& T \leq T_{max}. \\  \end{split}
\end{equation}
where $Q_p^k \in \mathbb{R}^2$ represents the coordinates of the centre of the formation. Additionally, \red{$OB$ is} the areas occupied by the obstacle cells, $L$ and $T$ denote the actual path length and coverage time, respectively. $v_{max}$ indicates the maximum  velocity of the mobile vehicle.  The nomenclature for the problem and our algorithm is summarised in Table \ref{tab:variables}.

\begin{table}[h]
 \centering
\caption {Nomenclature} 
\label{tab:variables}
  \centering
  \begin{tabular}{ll}
    \hline
     \textbf{Symbols} &  \textbf{Parameters} \\
      \hline
$L_{max}$ &  Path length budget \\
$T_{max}$ & Coverage time budget  \\
$CP_{max}$ &  Maximum coverage percentage \\
L & Actual path length  \\
T & Actual coverage time \\
CP &  Actual coverage percentage \\
MST &  Minimum spanning tree \\
$\hat{L}$ & Predicted path length \\
$\hat{T}$ & Predicted coverage time \\
$\hat{CP}$ & Predicted coverage percentage \\
$n_{q}$ & Number of mobile vehicles \\
$\theta$ & UGV heading \\
$q_p$ & UGV position in 2D \\
$Q_{p}$ & Centre position of the formation \\
CS & Grid cell size \\
BS & Block size \\
OB & Obstacle cells located in the UGVs’ total observation range \\
$\zeta^k$ & New cells covered by all UGVs' total observation area \\
$\sigma^k$ & Total observed cells at time $k$ \\
$l_o$ & Desired natural length \\
$l_a$ & Actual length \\
$F_{ij}$ & Virtual spring force vector between UGV $i$ and UGV $j$ \\
$D_{c_{ij},c_{i^{'}j^{'}}}$ & Connection direction from cell $c_{ij}$ to $c_{i^{'}j^{'}}$ \\
$F_g$ & Goal following force vector \\
$\omega_g$ & Target force weight \\
$\Delta$ & Blocked angles \\
\hline
  \end{tabular}
\end{table}

\subsection{Algorithm for Coverage Path Planning using Flexible Formation Control}

This section describes our algorithm for MCPP in a known environment. We begin with an informal description supported by diagrams here, then provide algorithmic details in the following sub-sections. The algorithm is designed to produce a coverage path that maximises the area of the environment that will be visited while meeting a given time budget. We use a binary and linear search to find the smallest grid cell size (CS) that will produce a coverage path in the presence of obstacles as per (\ref{eq:problem}). The algorithm proceeds as follows:
\begin{enumerate}
\item The algorithm starts by choosing a grid cell size and overlaying it on the given map. For example, grid lines can be seen in grey in Fig. \ref{fig:map_exp2}.
\item Next, a scanline algorithm is used to convert the environment geometry onto the grid as filled or free cells. The blue cells in Fig. \ref{fig:map_exp2} show (filled) obstacles, while yellow cells represent (free) open space.
\item The open space is then decomposed into variable-sized square blocks that follow the grid lines. These blocks are colored red, aqua, and purple in Fig. \ref{fig:map_exp2}.

\begin{figure}
	\begin{center}
		\begin{tabular}{cc}	
			\includegraphics[width=17.5pc]{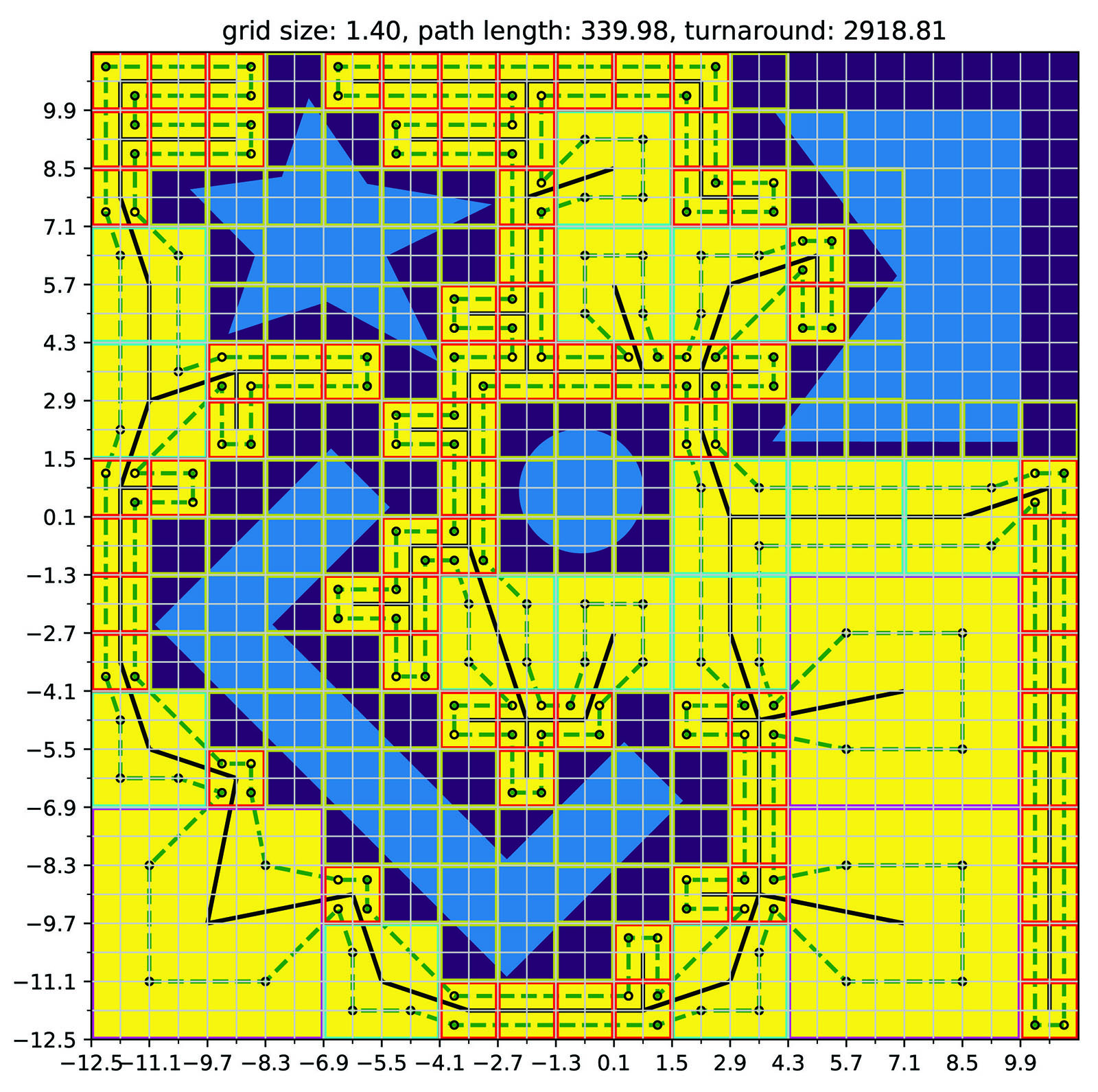} \\
			(a) \textit{}\\[6pt]
			\includegraphics[width=17.5pc]{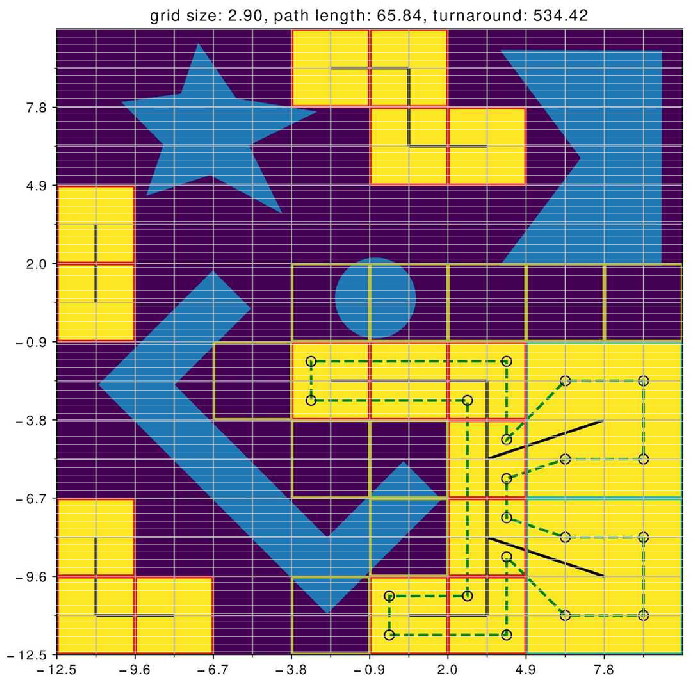} \\
			(b) \textit{} 
		\end{tabular}
		\caption{\footnotesize (a) demonstrates the coverage path obtained with a fine grid of 1.4m, resulting in a longer path; (b) presents an example with a coarse grid of 2.9m, leading to a shorter coverage path. The UGVs navigate through waypoints positioned at the bends of the path. The extent of the environment covered by the UGVs depends on the grid cell size due to the grid discretization. Blue polygons depict the obstacles, while purple cells indicate areas occupied by obstacles. Yellow cells represent the traversal areas, and purple cells with yellow lines depict the boundaries of obstacle areas. The small circles indicate sharp turns in the path. The connected green lines illustrate the robot path planned using the minimum spanning tree (MST) technique.}
		\label{fig:map_exp2}
	\end{center}
\end{figure}

% \begin{figure}
% 	\begin{center}
% 		\begin{tabular}{cc}	
% 			\includegraphics[width=17.5pc]{Figs/burger_big_arena_exp2} \\
% 			(a) \textit{Small grid cell size} \\[6pt]
% 			\includegraphics[width=17.5pc]{Figs/path_exp3}\\
% 			(b) \textit{Larger grid cell size} \\[6pt]
% 		\end{tabular}
% 		\caption{\footnotesize Examples of obstacle boundaries and predicted paths for different grid sizes.}
% 		\label{fig:map_big_exp1}
% 	\end{center}
% \end{figure}

\item The centre points of these blocks are joined by a minimum spanning tree, shown in black in Fig. \ref{fig:map_exp2}.
\item Afterward, a coverage path is constructed that      circumnavigates the spanning tree. This is shown as a dashed green line in Fig. \ref{fig:map_exp2}.
 \item Using the coverage path, the algorithm predicts how long coverage will take. Based on this predicted time and maximum path length, the binary search either outputs the final grid cell size or continues to dot point one above until a grid cell size is determined that will meet the budget. If the cell size increases above the observation range of the UGV, the algorithm will exit without a solution.
\item Once the cell size is chosen, waypoints are generated at each bend on the coverage path. These are shown as black circles in Fig. \ref{fig:map_exp2} and contain data about the size of the block they sit in.
\item The waypoints are then sent, in order, to a group of UGVs, which use the block size data to assume a formation that will enable them to pass through the block in a way that covers all the cells within it. The group of UGVs adapt their formation in response to both planned and real-time data from their sensors. Example formations are shown in Fig. \ref{fig:virtual_leader}.

\end{enumerate}

\begin{figure}
	\centering	\includegraphics[width=20.5pc]{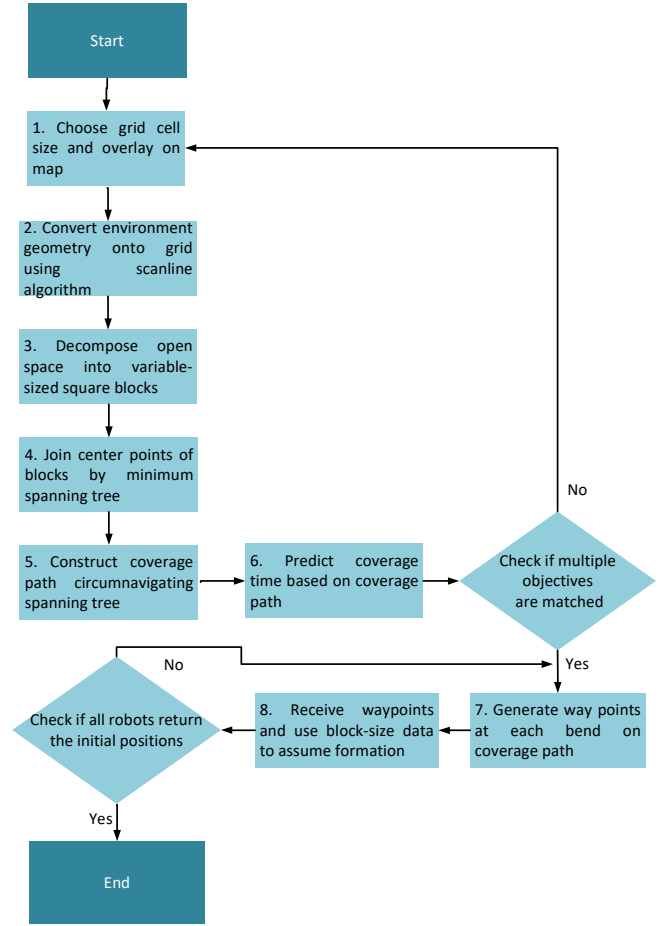}
	\caption{\footnotesize Flow chart of the whole strategy. See page 3 for details.}
	\label{fig:novel_method}
\end{figure}

Fig. \ref{fig:map_exp2}(a) has a fine grid, while Fig. \ref{fig:map_exp2}(b) has a coarse grid. The effect of the grid resolution is twofold. First, the obstacle boundaries (shown in dark purple cells) become coarser, blocking off (or de-prioritising) parts of the environment. For example, in Fig. \ref{fig:map_exp2} the area behind the large obstacle is blocked off. This, in turn, has the effect that the spanning tree becomes smaller, and the coverage path is shorter. These two effects contribute to the UGVs being able to meet a lower time budget by trading-off coverage percent. Further, Figure \ref{fig:novel_method} shows the flowchart of the algorithm, which provides a visual representation of the steps described in the algorithm list. We now consider the algorithm in detail.

\subsubsection{Interpreting Obstacle Geometry as a Grid}
We assume obstacles are input to the system as a set of vertices. A scanline algorithm \cite{clark1979} is used to classify grid cells as being either an obstacle or free space. The scanline algorithm scans through the grid horizontally. Locations of the intersection between the scan lines and the obstacle edges are detected. Each obstacle is then discretized into grid cells at all intersection points and between pairs of intersection points. 

\subsubsection{Creating Variable Sized Blocks}
Once the accessible areas are realized, a block-building algorithm is applied to group adjacent free grid cells into variable-sized blocks. The largest block size is chosen to accommodate the size of the group of UGVs when they are spread out. The remaining block sizes are calculated by progressively halving the largest size. 

We assume several window-based scans are performed from left to right and from bottom to top using each block size in turn. The largest block size is selected first, then gradually reduced to 1. If all grid cells in the scan window are free and do not belong to another block, they will be merged into a block and labelled with the block identifier. Otherwise, the scan window will skip and continue to the next position. After scanning all block sizes, a list of blocks covering the entire map is obtained. The block-building algorithm is summarised in Algorithm 1 of the Supplementary Material section.

\begin{figure}[H]
    \centering
    \includegraphics[scale=0.42]{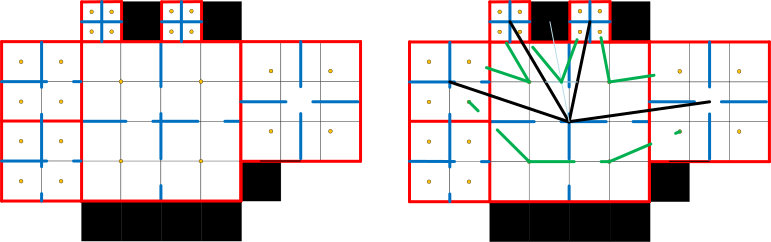}
	\caption{\footnotesize (left) Example of the neighbouring blocks in four directions and (right) Connections between neighbouring blocks. The white and black cells are free and obstacle cells, respectively. The red lines are the boundary lines of the block. Blue dash lines and yellow dots are the boundary lines and the center of block parts.}
    \label{fig:block_example}
\end{figure}

Next, the connectivity between the blocks needs to be established. Every grid cell always has four connected neighbours (top, bottom, left, right). See Fig. \ref{fig:block_example} for an example. The white and black cells are free and obstacle cells, respectively. The black lines are the edges of the spanning tree. The green dash lines are the generated paths. Blue dash lines and yellow dots are the boundary lines and the center of block parts. If the two adjacent cells belong to two different blocks, these two blocks will be marked as connected.  Meanwhile, the connection direction of the two blocks is equal to the connection direction of the two cells. However, there is also no connecting edge between the center block and the bottom left block in the MST (black line) because the top left block is closer to the bottom left block than the center block. The connection direction $d \in \{TOP, LEFT, BOTTOM, RIGHT\}$, from cell $c_{ij}$ to cell $c_{i^{'}j^{'}}$ $d_{c_{ij},c_{i^{'}j^{'}}}$ is derived as:
\begin{equation}
d_{c_{ij},c_{i^{'}j^{'}}} = \begin{cases}
LEFT~,~[i-i^{'},j-j^{'}]=[-1,0], \\
RIGHT~,~[i-i^{'},j-j^{'}]=[1,0], \\
BOTTOM~,~[i-i^{'},j-j^{'}]=[0,-1], \\
TOP~,~[i-i^{'},j-j^{'}]=[0,1]. \\
\end{cases}
\end{equation}

The block to which grid cell $c$ belongs is $b_c$. Assume blocks of adjacent cells $c_{ij}$ and $c_{i^{'}j^{'}}$ are different. The direction from the block of cell $c_{ij}$ to the block of cell $c_{i^{'}j^{'}}$ is given by:
\begin{equation}
d_{b_{c_{ij}},b_{c_{i^{'}j^{'}}}}=d_{c_{ij},c_{i^{'}j^{'}}}
\end{equation}

After building a graph with the connection between blocks, the minimum spanning tree algorithm will be used to find a spanning tree. Connections that do not belong to the spanning tree will be removed. 

\subsubsection{Minimum Spanning Tree}
A spanning tree is a subset of an undirected graph $G(V,E) $ that connects all the vertices $V$ of the graph with a minimum number of edges $E$. A spanning tree cannot contain cycles or disconnected nodes. However, a connected and undirected graph $G$ may be connected with more than one spanning tree since every vertex can be connected from many directions. The cost of the spanning tree is the sum of edge weights in the tree. 

The most efficient spanning tree can be found by a minimum spanning tree algorithm. The MST algorithm constructs a tree including every vertex, where the sum of the weights of all the edges in the tree is minimized. Prim's algorithm \cite{greenberg1998} is used to construct the spanning tree by computing an edge with the least weight and adding it to the growing spanning tree. Prim's algorithm is selected for our work since it is one of the fastest algorithms in dense graphs \cite{huang2009}. 

\subsubsection{Coverage Path Planning Algorithm}

% \begin{figure}[h]
%     \centering
%     \includegraphics[scale=0.35]{Figs/block_example3}
% 	\caption{\footnotesize The schematic diagram of four parts of a 3$\times$3 cells block. The red lines are the boundary lines of the block.}
%     \label{fig:block_example3}
% \end{figure}

\begin{figure}
	\centering
	\includegraphics[width=19pc]{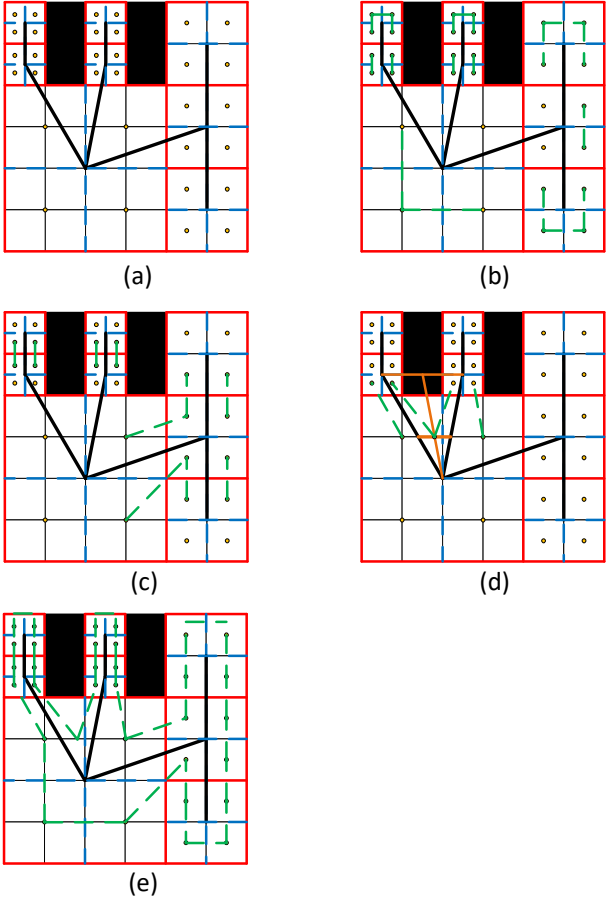}
	\caption{\footnotesize (a) illustrates the MST itself, while (b-d) depict three specific cases: (b) No adjacent block, (c) One adjacent block, and (d) Two or more adjacent blocks. (e) shows a combination of all three cases (b), (c), and (d). In the figure, white cells represent free cells, while black cells represent obstacle cells. The black lines indicate the edges of the spanning tree, while the green dashed lines represent the generated paths. The orange lines illustrate the addition of joint points according to lines 16-20 of Algorithm 2.}
	\label{fig:block_example2}
\end{figure}

For the purpose of path planning, each block is logically divided into four parts, as depicted in Fig. \ref{fig:block_example}, and the center of each part serves as the connection point along the constructed route. The construction of the coverage path depends on the number of adjacent blocks in a given direction, which can be categorized into three cases: (1) no adjacent block, (2) one adjacent block, or (3) multiple adjacent blocks. In the first case, the centers of the two parts are simply connected. In the second case, if the adjacent block is already connected to the current block, it is ignored. Otherwise, the two parts of the current block are connected to the adjacent parts of the neighboring block in the same direction. In the third case, the adjacent blocks are sorted in the same direction, and for each adjacent pair of blocks, a joint point is generated as the intersection point between two lines. The first line connects the midpoint between the two centers of the adjacent blocks to the center of the current block, and the second line connects the center of the two adjacent parts of the current block. Fig. \ref{fig:block_example2} (d) illustrates these lines, and the relevant formulas are located in lines 16-20 of Algorithm \ref{alg:pp_algo}. Finally, the adjacent parts of the adjacent blocks are connected to the generated joint points. Figs. \ref{fig:block_example} and \ref{fig:block_example2} visually demonstrate the linking of blocks and the generation of the UGV route.

Denote $D_{d,b}$ as the list of neighbor blocks of the block $b$ in the direction $d$, $Q_{c,b}$ as the coordinate of the center of the part $c \in \{TL, TR, BL, BR\}$ of the block $b$, $P_{C,b}$ as the coordinate of the center of the block $b$, $P_{L,b}$ as the coordinate of the left edge of the block $b$, $N_b$ as the list of neighbor blocks of the block $b$. The specific implementation is given in Algorithms 1 and \ref{alg:pp_algo}.

\begin{algorithm}
\setcounter{algorithm}{1}
\caption{Path Planning Algorithm}
\label{alg:pp_algo}
\begin{algorithmic}[1]
\State \textbf{Inputs}: List of blocks $B_k$, spanning tree $ST = \{E_{B_i,B_j} | i \neq j, i,j=1,\dots,N_B\}$, starting point $p_{start}$
\State \textbf{Outputs}: List of path elements $P$.
\For{$b \in B_k$}
    \State $b_{LEFT} \gets D_{LEFT,b}$
    \If{$|b_{LEFT}|>0$}
        \If{$|b_{LEFT}|=1$}
            \If{$|D_{RIGHT,b^1_{LEFT}}|=1$}
                \State Add path $(Q_{TL, b}, Q_{TR, b^1_{LEFT}})$ to $P$
                \State Add path $(Q_{BL, b}, Q_{BR, b^1_{LEFT}})$ to $P$
            \EndIf
        \Else
            \State Sort $b_{LEFT}$ by $x$ in ascending order
            \State $n \gets |b_{LEFT}|$
            \State Add path $(Q_{TL, b}, Q_{TR, b^1_{LEFT}})$ to $P$
            \For{$i=1,\dots,n-1$}
                \State $middle \gets \frac{P_{C,b^i_{LEFT}}+P_{C,b^{i+1}_{LEFT}}}{2}$
                \State $center \gets P_{C,b}$
                \State $joint_x \gets P_{L,b}+\frac{BS_b}{4}$
                \State $joint_y \gets middle_y + \frac{(joint_x, middle_x)*(center_y-middle_y)}{center_x-middle_y}$
                \State $joint \gets (joint_x, joint_y)$
                \State Add path $(Q_{BR,b^i_{LEFT}},joint)$ to $P$
                \State Add path $(joint,Q_{TR,b^{i+1}_{LEFT}})$ to $P$
            \EndFor
            \State Add path $(Q_{BL, b}, Q_{BR, b^n_{LEFT}})$ to $P$
        \EndIf
    \Else
        \State Add path $(Q_{TL, b}, Q_{BL, b})$ to $P$
    \EndIf
    \State Do the same for the RIGHT, TOP, BOTTOM directions
    \State Sort $P$ in order of proximity to starting point $p_{start}$
\EndFor
        
\end{algorithmic}
\end{algorithm}

\subsubsection{Predicting Time to Follow the Path}
The turnaround time $\hat{T}$ and total angle difference $\hat{\theta}$ are estimated by:
\begin{equation}
\begin{split}
% & \hat{L} = \sum_{i=1}^{|P|}{|p_i|}, \\
& \hat{\theta} = \sum_{i=1}^{|P|}{|\alpha_{p_i}-\alpha_{p_{i+1}}|} \\
& \hat{T} = \frac{L}{v} + \frac{n_t*\hat{\theta}}{\omega}, \\
% & \hat{CP} = \frac{\sigma^T}{C_{free}+OB}100\%, 
\end{split}
\end{equation}
where $\alpha_{p_i}$ is the orientation of the $i^{th}$ line $p$ with respect to the x-axis. The component  $\frac{n_t*\hat{\theta}}{\omega}$ in the $\hat{T}$ equation is the leader's rotational time. The total path length $\hat{L}$ and the coverage percentage $\hat{CP}$ are defined as in (\ref{eq:problem}). At sharp corners, the leader only rotates around its heading; whereas, the followers modify their positions. The formation's rotational motion at the center of the formation, therefore, can be approximated as that at the leader position.

\subsubsection{Optimising the Grid Cell Size to Meet the Time Budget}

We use a binary search strategy with low memory and low complexity \cite{huang2012new} to identify the appropriate grid cell size that produces a path length that can be followed within a given time budget. Binary search repeatedly divides the search interval in half of the lower $\underline{m}$ and the upper $\overline{m}$ bounds. If all estimated values are equal to the given metrics, the search is completed. The searching also stops when the lower $\underline{m}$ is greater than or equal to the upper $\overline{m}$. Otherwise, if all estimated values of the current CS are more than the given metrics, the search narrows the interval in the upper half. Otherwise, the search recurs in the lower half. However, the maximum coverage percentage metric cannot be calculated accurately due to intermittent scanning values.

According to Algorithm \ref{alg:bl_algo}, the binary search begins to compute $\hat{L}$ and $\hat{T}$ for every chosen grid cell size. It repeatedly checks until the two target values, which are less than $\hat{L}$ and $\hat{T}$, are obtained. Then, the linear search resumes estimating $\hat{CP}$ for larger subsequent cell sizes until the final objective (maximum coverage percentage) is attained within 10 next cell size values by the linear search optimisation algorithm.

\begin{algorithm}
\caption{Grid Cell Size Optimisation Algorithm}
\label{alg:bl_algo}
\begin{algorithmic}[1]
\State \textbf{Algorithm Parameters}: $\hat{L}$, $\hat{T}$, and $\hat{CP}$.
\State \textbf{Inputs}:  CSs, $\delta$.
\State \textbf{Outputs}: $\bar{CS}$.
\State Consider a cell size list with $n_c$ element, $CS_0,..., CS_{n_c-1}$, and three target metrics $\delta$. $m$ is the pointer's position while $\underline{m}$ and $\overline{m}$ are the left and right positions of search area.
\State Initialize $\underline{m} \gets 0$, $\overline{m} \gets n_c-1$, $i \gets 0$.
\While{$\underline{m}<\overline{m}$}
    \State Compute $\hat{L}_m$ and $\hat{T}_m$ at the middle position $m = \frac{\underline{m}+\overline{m}}{2}$.
    \If{$\hat{L}_m > L_{max}$ and $\hat{T}_m > T_{max}$}
        \State Set $\underline{m} \gets m+1$
    \ElsIf{$\hat{L}_m = L_{max}$ and $\hat{T}_m = T_{max}$}
        \State Set $\underline{m},\overline{m} \gets m$
        \State \textbf{break}
    \Else
        \State Set $\overline{m} \gets m-1$
    \EndIf
\EndWhile
\State $m \gets \min(\underline{m}, \overline{m})$
\While{$m<n_c$}
    \State Compute $\hat{CP}_{m}$ at the $m$ position.
    \If{$\hat{CP}_{m} > CP_{max}$}
        \State $CP_{max} \gets \hat{CP}_{m}$
        \State $i \gets 0$
    \Else
        \State $i \gets i + 1$
    \EndIf
    \If{$i = 10$}
        \State $\textbf{break}$
    \EndIf
    \State $m \gets m + 1$
\EndWhile
\State Return $m$ and $CS_m$
\end{algorithmic}
\end{algorithm}

\subsubsection{Formation Control}
\begin{figure}
	\centering
	\begin{subfigure}[b]{0.23\textwidth}
		\includegraphics[width=\textwidth]{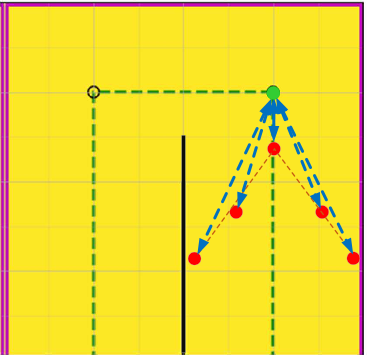}
		\caption{\footnotesize \textit{V-shaped formation.}}
		\label{fig:v_shaped}
	\end{subfigure}
	\begin{subfigure}[b]{0.2\textwidth}
		\includegraphics[width=\textwidth]{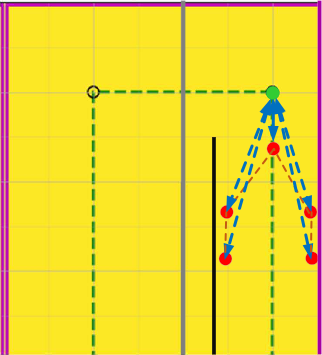}
		\caption{\footnotesize \textit{U-shaped formation.}}
		\label{fig:u_shaped}
	\end{subfigure}
	% 	\begin{subfigure}[b]{0.2\textwidth}
	% 	\includegraphics[width=\textwidth]{Figs/que_formation.eps}
	% 	\caption{\textit{Line-shaped formation.}}
	% 	\label{fig:que_shaped}
	% \end{subfigure}
    \caption{\footnotesize V- and U-formations obtained by a dynamic role assignment. UGVs maintain their distance from a `virtual leader' to maintain their formation. Red circles represent the follower robots, while green circles indicate the virtual leader. Black circles are the cell centre. Dashed green lines and black lines are the planned trajectory and the spanning tree, respectively.}
    \label{fig:virtual_leader}
\end{figure}

Each of the $n_q$ UGVs is given an identification number subscript $q_0$, $q_1$, etc. Each real UGV is assumed to be connected by a virtual spring system to its virtual leader, representing the physical interconnection between the UGVs and wireless communication. Referring to Fig. \ref{fig:virtual_leader}, orange lines represent the desired formation. Red circles indicate the follower UGVs, while green circles illustrate the virtual leader. Black circles are  waypoints. Dashed green lines show the planned path. The black line is the spanning tree. Multiple UGVs ($q_1,..., and~q_5$) will be controlled to move in a formation whose pattern depends on the various block sizes. $q_0$ is treated as the virtual leader; the remaining UGVs are followers. Once virtual UGV $q_0$ and UGV $q_i$ ($i \geq 1$) are linked together through virtual springs (VSs), spring forces are generated between them. 

Based on the desired natural length vector $l_o$ and actual length vector $l_a$ of each VS, the common control rules of the VS method can be set.

The degree of difficulty in assigning a specific character for a UGV in the formation is defined as a character cost. The cost function equation is expressed as:
\begin{equation}
\phi_{ij} = \omega_{cx} |l_{a_{0ix}}| + \omega_{cy} |l_{a_{0iy}}| + \omega_{c\theta} |l_{a_{0i\theta}}|, \forall i, j \geq 1
\end{equation}
where $\phi_{ij}$ is the cost of the UGV $i$ to obtain the $j^{th}$ goal role. $l_{a_{0ix}}$ and $l_{a_{0iy}}$ are the relative positions from the actual UGV position to the virtual leader position along the $x$ and $y$ axes. $l_{a_{0i\theta}}$ is the angular difference between the initial direction and the target direction. $\omega_{cx}$, $\omega_{cy}$, and $\omega_{c\theta}$ are weight constants.

% Define a character set matrix $S = [S_1 S_2 ... S_{n_q}]$.

After all costs for each character in the formation are calculated, the UGV with the maximum cost will be assigned a corresponding role in $fo$, as shown in Fig. \ref{fig:virtual_leader}(a) and the following equation:
\begin{equation}
fo_j = i~,~if~\phi_{ij} > \phi_{hj},\forall i \neq h,~i~\&~h \leq n_q, \forall fo_k > i, k \neq j. 
\end{equation}

A relative position-based formation control method is derived here to maintain the desired formation shape for networked UGVs. The natural lengths of the springs $l_0 = \{l_{01},...,l_{0n_q} \}$ are set according to the geometrical requirements for the desired formations. 

Depending on the block size (BS), there are three formation shapes: the first is V-shaped (V-formation), the second is the U-shaped U-formation, and the third is a queuing (line) formation (Q-formation). The V-formation or U-formation is selected when the BS is greater than one. For a block size of 1, the Q-formation is used. Using the BS information and the formation type, the desired relative positions for each follower with respect to the virtual leader are computed to generate an obstacle avoidance formation, as described in Algorithm 4 of the Supplementary Material section. The control input to each ground vehicle is the resultant of the force vector generated by the virtual spring pairs connected to the UGVs.

%%The binary condition to switch the formations is as follows:
%\begin{equation}\label{eq:formation_sw}
%\begin{split}
%&if BS > 1, VS = 1 \land QS = 0, \\
%&if BS = 1, VS = 0 \land QS = 1.  
%\end{split}
%\end{equation}

\subsubsection{Path Tracking Algorithm}
After the leader UGV's coverage trajectory is computed, an online path planner outputs a series of way points $wp = {wp_0,..., wp_k,..., wp_{n_w}}$ around the trajectory for the UGV navigation, where $n_w$ denotes the number of way points. Further, each UGV in the formation knows the remaining UGVs' position information.

Using the virtual leader strategy, the spanning-tree path is shared among UGVs after the calculation process is completed. Based on the virtual vehicle tracking error of follower 1 (the closest follower, named $fo_1 \in fo$), the proposed virtual velocity for the virtual vehicle $v_{q_0}$ is:
\begin{equation}
v_{q_0} = v \bigg(1-\min\big(\max(\frac{|{l_{a_{01}}}|-\underline{l_{q_{0}}}}{\overline{l_{q_{0}}}-\underline{l_{q_{0}}}},0),1\big)\bigg),
\end{equation}
where $|{l_{a_{01}}}|$ represents the distance between the virtual leader and follower 1, $\overline{l_{q_{0}}}, \underline{l_{q_{0}}}$ represent the maximum and minimum distance between the virtual leader and follower 1 to reach the minimum and maximum speed, respectively.

After every time step, the next way point $wp_{k+1}$ is produced:
\begin{equation}
wp_{k+1} := wp_{k}+v_{q_0}~dt,
\end{equation}
where $dt$ is the sample time.

Based on the Euler distance between the leader and the goal, the goal following force vector $F_g$ acting on the leader is derived as:
\begin{equation}
F_g = \omega_g*(wp_{k+1}-q_{p_0}),
\end{equation}
where $\omega_g$ is the attractive force weight.

The proposed formation control approach is distributed to each UGV to guarantee that if any UGV fails the rest can adopt new roles and continue to track the virtual leader.

\newcommand{\floor}[1]{\left\lfloor #1 \right\rfloor}
\newcommand{\ceil}[1]{\left\lceil #1 \right\rceil}
\newcommand{\atantwo}{\operatorname{atan2}}

\subsubsection{Closest-Safe-Angle-Based Obstacle Avoidance}
In order to prevent further collisions, a LiDAR sensor system is equipped on each UGV to measure distances ($d_o$) and angles ($\alpha$) from any surfaces. It works by emitting pulsed light waves in all directions and measuring how long it takes for them to bounce back off surrounding objects. To be convenient for further computation, the LiDAR information chain is converted to Cartesian coordinates as follows:
\begin{equation}
\alpha_\iota = \iota~\phi + \theta, \forall~0\degree \leq \alpha_\iota \leq 360\degree,
\end{equation}
\begin{equation}
p^\iota_o = (x^\iota_o,y^\iota_o) = (d^\iota_o~\cos{\alpha_\iota},d^\iota_o~\sin{\alpha_\iota}),
\end{equation}
where $\phi$ denotes an angular distance between measurements while $\theta$ illustrates the UGV heading and $\iota$ stands for the measurement step. $p^\iota_o = (x^\iota_o,y^\iota_o)$ is the obstacle position at the $\iota^{th}$ scan time. Additionally, $d^\iota_o$ is the $\iota^{th}$ relative distance measured at the $\iota^{th}$ $\alpha$ scanning angle.

When any of the UGVs or obstacles move or violate the avoidance radius $R_av$, the UGV's angular width causing possible collisions called the blocked angles $\alpha_b$, can be estimated and incorporated into the planner. To guarantee safe passage, the UGV is steered towards angles that are not blocked, called safe angles $\alpha_s$.

A discrete list of all possible heading angles between 0 and 2$\pi$ is generated based on the angle increment $\phi$. For example, if $\phi = \frac{\pi}{180}$, the list's size $n$ will be 360. The algorithm then searches for all safe angles in the list: $0 \leq \alpha^\iota_s < 2 \pi, \alpha^\iota_s \in \alpha \land 1 \leq \iota \leq n$.

\begin{figure}[H]
    \centering
    \includegraphics[scale=0.8]{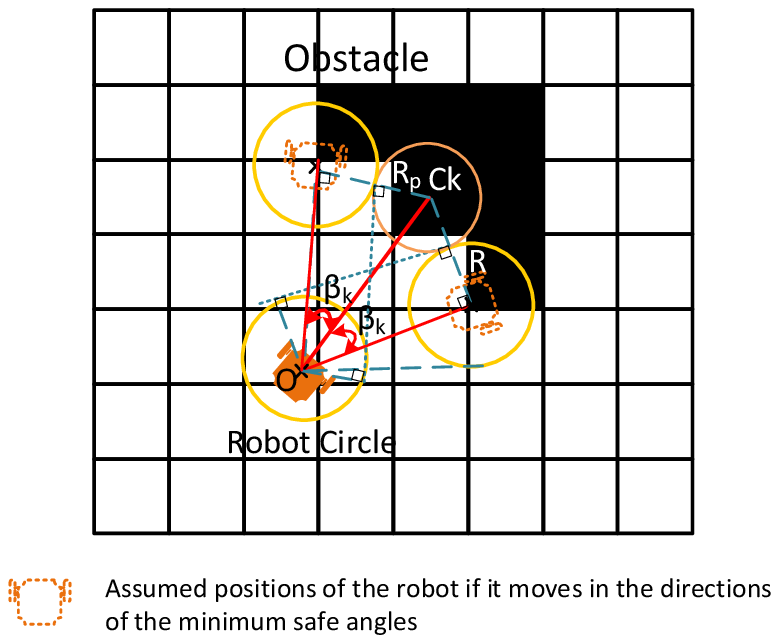}
    \caption{\footnotesize A demonstration of avoidance strategy.}
    \label{fig:safe_angle}
\end{figure}

We define a collection of obstacle cells that need to be avoided: $P' = {P^k_o: 1 \leq k \leq m}$, the UGV circle's radius $R$, the UGV centre $O$, the obstacle circle's radius $R_p$, the $P^k_o$ centre $C_k$, and the relative angle $\eta^k$ between the centre line $OC_k$ and the $X-Axis$. The two tangent lines between the circle of the obstacle $P^k_o$ and the UGV circle are constructed as shown in Fig. \ref{fig:safe_angle}. Next, a set of two symmetric $k^{th}$ blocked angles [$-\beta^k$,$\beta^k$] with respect to the centre line $OC_k$ can be expressed as:
\begin{equation}
[-\beta^k;\beta^k] = [-\arcsin{\frac{R+R_p}{OC_k}},\arcsin{\frac{R+R_p}{OC_k}}].
\end{equation}

All global heading angles $\alpha$ located in the $\beta$ angle's width are treated as blocked angles for the relevant UGV to transverse. The collection of angles blocked by the obstacle $P^k_o$, named $\Delta_k$, can be defined as:
\begin{equation}
\Delta_k:\{\alpha_\iota: \leq \alpha_\iota \leq \eta^k+\beta^k,\alpha_\iota \in \alpha\}.  
\end{equation}

A complete list of the blocked angles $\Delta$ is obtained from:
\begin{equation}
\Delta = \cup^{k=1}_{m} \Delta_k.
\end{equation}
%Recapitulating all sets $D_k$

Now, a list of safe angles $S$ can be established by excluding the list $D$ from the list $\alpha$:
\begin{equation}
S = \alpha - \Delta.
\end{equation}

If $S = \emptyset$, the UGV would halt ( $V = 0$ ) until any safe angle is scanned. Otherwise, the safe angle closest to the target angle $\alpha_t$ is selected as follows:
\begin{equation}
\alpha_r = min(|\alpha_\iota - \alpha_t|), \alpha_\iota \in S. 
\end{equation}

The outcome of our obstacle avoidance strategy is the UGV's desired minimum orientation $\alpha_r$. If no obstacles are predicted, the virtual force vector $(v_x, v_y)$ is fused from all force components (target force and spring force). Otherwise, this vector is turned towards the angle $\alpha_r$ in order to generate a new avoiding vector $v_{av}$ as in (\ref{eq:fused_force}).
\begin{equation}\label{eq:fused_force}
v_{av} = \begin{bmatrix}
v_{avx} \\
v_{avy}
\end{bmatrix}^T = 
\begin{bmatrix}
v_x cos(\alpha_r) - v_y sin(\alpha_r) \\
v_x sin(\alpha_r) + v_y cos(\alpha_r)
\end{bmatrix}^T,
\end{equation}

Based on (\ref{eq:fused_force}), the control input of each UGV is computed as:
\begin{equation}
u = \begin{bmatrix}
v \\
\omega
\end{bmatrix} = \begin{bmatrix}
v = |v_{av}|, \\
\omega = atan.(v_{av}),
\end{bmatrix}
\end{equation}

\subsection{Complexity Analysis}
This section discusses the time complexity of the algorithm components mentioned above. Dynamic formation implementations demonstrate a worst-case time complexity of $O(n_q^2)$ where $n_q$ stands for the number of mobile UGVs. This is because each physical UGV must inspect its position relative to the virtual leader and then transfer the relative position information to every other UGV to determine its role in the formation. The other significant element of our approach is the exchange of coverage matrices between UGVs. On receipt of such data from another UGV, each UGV must compare $C_W \times C_H$ cells to update its coverage and obstacle matrices. This process delivers time complexity $O(C_WC_H)$. Hence, the worst-case time complexity would be $O(n_q^2) + O(C_WC_H)$ if all agents exchange their coverage matrices and their own relative position information with all other agents. The algorithm is theoretically scalable to large formations and environments due to no exponential terms. In our case where there is a small number of UGVs, $O(C_WC_H)$ is the dominant term.

\section{Experiments}

This section presents a range of experiments designed to evaluate the performance of our planning and prediction engine in various scenarios. In Section IV.A, we demonstrate the engine's accuracy in predicting execution time by conducting an experiment with a simulated UGV. This experiment shows that the engine can efficiently find a suitable coverage path plan that meets a given time budget in only a short time. 

In Section IV.B, we conduct a series of comparative experiments using our novel coverage path planning with formation control (CPPF), the coverage path planning with swarming, and the frontier-led swarming \cite{tran2022frontier} on simulated Jackal UGVs. These experiments investigate the impact of UGV speed and time budget on path following performance while maintaining the group and order of the UGV formation. Section IV.C compares the multi-robot coverage path planning performance of CPPF and another method that uses a quadtree data structure without physical limits \cite{huang2020}. This comparison provides insights into the strengths and weaknesses of each approach to the optimal coverage path planning problem in cluttered environments. In the simulated experiments, we used the same settings (number of agents, maximum exploration time, path length, environmental setups, and initial locations) to guarantee a fair comparison between algorithms.

In Section IV.D, we describe experiments conducted on real Jackal UGVs in outdoor environments. These experiments aim to evaluate the engine's performance in real-world conditions. Finally, we summarize the experiments in Section IV.E, providing an overview of the findings and their implications. 

\subsection{Experiment 1: Characterising Prediction Performance}
\subsubsection{Experiment 1 Setup} 
To verify the effectiveness of the prediction engine, a simulated, cluttered map of $25m\times25m$ is used  (see Fig. \ref{fig:map_exp2}). Four static and complex-shaped obstacles (e.g., star, U-shaped, cylinder, and castle-shaped) were placed randomly. Cell sizes ranging from 0.75m to 3m are used to generate coverage paths and time-to-follow predictions. A single simulated UGV was then permitted to follow the path and the time taken compared to the prediction. 

A single small Turtlebot UGV was used in these experiments so that we could examine the approximate path following performance without the complexity of formation control. The UGV was equipped with a 360-degree LiDAR sensor for reactive obstacle avoidance while path following. The forward speed $v$ of the UGV was initialised to 0.14$m/s$. The maximum turn rate when maneuvering was set to $0.7rad/s$. Other parameters of this experiment are summarised in Table \ref{tab:pred_Expt}. The i7-1260P CPU, Ubuntu Focal (20.04), ROS Noetic framework, and Python 3.8 are used to program and implement the planning approaches. The dynamic behavior of all vehicles was simulated in Gazebo. The sampling time of the whole system is set at 30Hz.

% \begin{figure}
% 	\begin{center}
% 		\begin{tabular}{cc}	
% 			\includegraphics[width=15.5pc]{Figs/big_area} \\
% 		\end{tabular}
% 		\caption{\footnotesize Experiment 1 environment for testing prediction  performance. Cell sizes range from 0.75m to 3m.}
% 		\label{fig:env_set}
% 	\end{center}
% \end{figure}

\begin{table}[h]
 \centering
\caption {Experiment 1 parameters and their experimental values} 
\label{tab:pred_Expt}
  \centering
  \begin{tabular}{lll}
    \hline
     \textbf{Parameter} &  \textbf{Description} &  \textbf{Value} \\
      \hline
$\delta_d$ &  Distance tolerance & 0.05m  \\
$\delta_\theta$ &  Angle tolerance & 15\degree  \\
$\omega_{g}$ &  Target force weight & 1.1 \\
\hline
  \end{tabular}
\end{table}

\subsubsection{Experiment 1 Performance Metrics}
Three criteria are chosen to evaluate the prediction performance of the proposed algorithms. They are: 
\begin{itemize}
    \item PLD: Difference between the predicted and actual coverage path length,
    \item TTD: Difference between the predicted and actual turnaround time (where turnaround time is defined as the time to complete following the path), 
    \item CPD: Difference between the predicted and actual coverage percentage. 
\end{itemize}

\subsubsection{Experiment 1 Prediction results} 
Tables \ref{tab:pld_metrics}-\ref{tab:tt_metrics} show the results from the prediction module versus the Gazebo path following simulation. PLD was under 25m and TTD under 300s. Our observation of the UGV's movement revealed that the need to slow down to perform turns was the main factor causing the error between predicted and actual performance. 

\begin{figure}
	\begin{center}
		\begin{tabular}{cc}	
			\includegraphics[width=21.5pc]{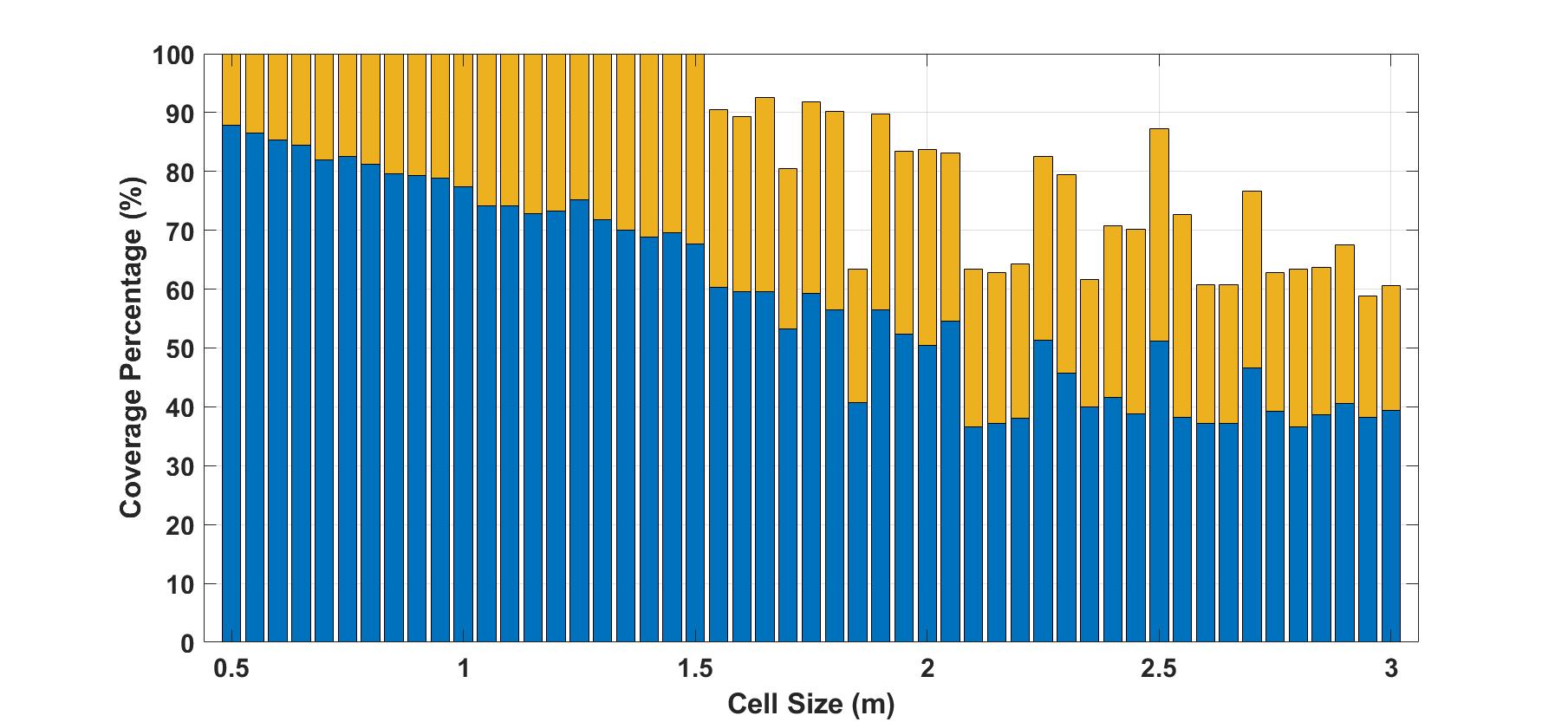} \\
			(a) \textit{Predicted coverage percentage.}\\[6pt]
			\includegraphics[width=21.5pc]{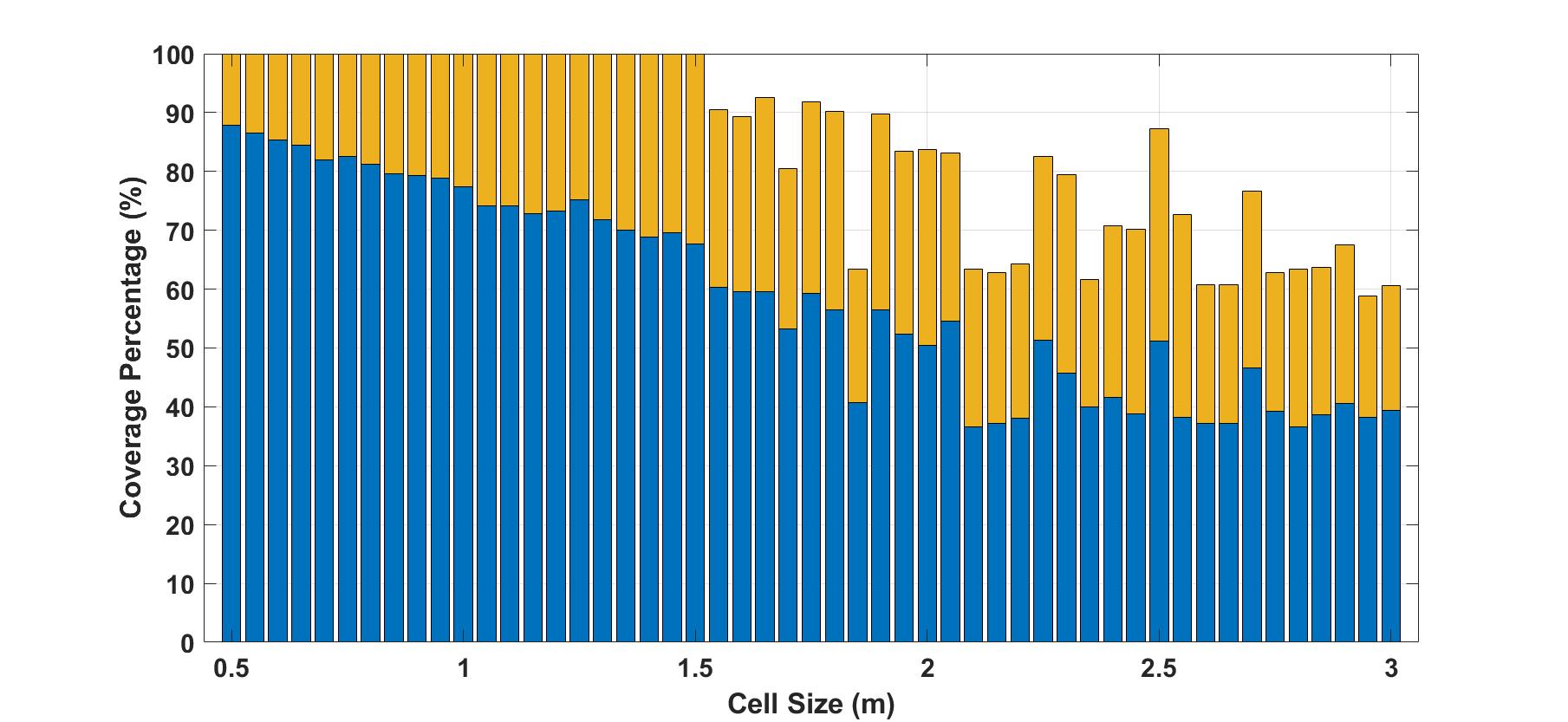} \\
			(b) \textit{Actual coverage percentage.} 
		\end{tabular}
		\caption{\footnotesize Experiment 1: Predicted and actual coverage percent metrics for different grid cell sizes. Blue bars represent the direct coverage made by the UGV rolling through a cell, while yellow bars indicate the indirect coverage achieved by the UGV's LiDAR sensor.}
		\label{fig:cp_metrics}
	\end{center}
\end{figure}

Fig. \ref{fig:map_exp2} shows two examples of the predicted paths for different grid cell sizes. We can see that the larger grid cell size causes a coarser boundary around obstacles (shown in purple around the blue obstacles). Fig. \ref{fig:cp_metrics} shows that this has the effect of reducing the total area that will be covered by the UGVs when the grid cell size is very large. Thus, while CPD was 0\%, the actual coverage percent reduces with increasing grid cell size. This has the effect of meeting a tighter budget. 

We also observed that the lower the grid cell size, the greater the computation time to make a prediction.  For example, the computation time for the 0.75m cell size is approximately 6s, while that for cell size above 1.5m does not go above 1.5s.

\begin{table}[h]  
\caption{\footnotesize Experiment 1: Predicted and actual path length (PPL and APL) metrics for different grid cell sizes.}\label{tab:pld_metrics}% title name of the table  
\centering % centering table  
\begin{tabular}{l c r r } % creating 10 columns  
\hline\hline   
Cell Size (m) &  PPL (m) & APL (m) & PLD (m)
\\ [0.5ex]  
\hline   
% Entering 1st row  
0.75 & 1228.5 & 1223.87 & 4.63  \\ 
1.0  & 890.0 & 882.37  & 7.63 \\
1.25 & 712.5 & 689.05 & 23.45 \\         
1.5 & 501 & 490.33 &  10.67 \\
1.75 & 378 & 369.98 &  8.03 \\ 
2.0 & 224 & 215.46 &  8.54 \\ 
2.25 & 252 & 245.77 & 6.23 \\ 
2.5 & 220 & 214.75 &  5.25 \\ 
2.75 & 110 & 106.34 & 3.66 \\
3.0 & 78 & 75.47 &  2.53 \\ 
\hline 
\end{tabular}  
\end{table}

\begin{table}[h]  
\caption{\footnotesize Experiment 1: Predicted and actual turnaround time (PTT and ATT) metrics for different grid cell sizes.}\label{tab:tt_metrics}% title name of the table  
\centering % centering table  
\begin{tabular}{l c r r } % creating 10 columns  
\hline\hline   
Cell Size (m) &  PTT (s) & ATT (s) & PLD (s)
\\ [0.5ex]  
\hline   
% Entering 1st row  
0.75 & 12114.06 & 12161.37 & -47.3  \\ 
1.0  & 8640.02 & 8339.27  & 300.75 \\
1.25 & 6440.17 & 6284.14 & 156.03 \\         
1.5 & 4386.41 & 4284.1 &  102.31 \\
1.75 & 3180.21 & 3183.306 & -3.09 \\ 
2.0 & 1914.16 & 1846.89 &  67.27 \\ 
2.25 & 2037.86 & 2019.13 & 18.74 \\ 
2.5 & 1777.88 & 1756.58 &  21.3 \\ 
2.75 & 884.45 & 876.92 & 7.53 \\
3.0 & 597.53 & 626.17 & -28.63 \\ 
\hline 
\end{tabular}  
\end{table}

\subsection{Experiments 2-4: Characterising Path Following Performance}

In this section, we study the efficacy of our formation control system (denoted CPPF) in following a planned coverage path. We examine the impact of the top speed of the UGVs and the time budget on performance. We include three comparative algorithms: the first (denoted CPPS and shown in Algorithm \ref{alg:adaptive_swarm}) substitutes the leader-follower flexible formation strategy with a rule-based swarm strategy. A swarm can adapt its shape reactively to wider or narrower parts of the environment but does not require a leader. The second comparative approach (denoted FS \cite{tran2022frontier}) uses a reactive, frontier-led swarming strategy with no planning. 

\begin{algorithm}
\setcounter{algorithm}{4}
\caption{Swarming waypoint follower}
\label{alg:adaptive_swarm}
\begin{algorithmic}[1]
\State \textbf{Algorithm Parameters}: alignment radius scale factor $\mathcal{K}_a$, cohesion radius scale factor $\mathcal{K}_c$, separation radius scale factor $\mathcal{K}_s$.
\State \textbf{Inputs}: cell size $CS$, block size $BS$, number of UGVs $n_q$
\State \textbf{Outputs}: swarm force $\sigma$. 
\State The alignment radius $R_{a}$, cohesion radius $R_{c}$, separation radius $R_{s}$ are determined as follows:
\begin{equation}
\begin{split}
R_{a} &= \mathcal{K}_a\frac{CS~BS}{n_q} \\
R_{c} &= \mathcal{K}_c\frac{CS~BS}{n_q} \\
R_{s} &= \mathcal{K}_s\frac{CS~BS}{n_q}
\end{split}
\end{equation}
\State The swarm force $\sigma$ is computed according to \cite{tran2022frontier} with $R_{a}$, $R_{c}$, $R_{s}$
\end{algorithmic}
\end{algorithm}

\subsubsection{Experiments 2-4 Simulation Setup}
The simulation environments in these experiments are implemented on the same PC running Ubuntu, ROS Noetic, and Gazebo's Jackal models as in Experiment 1. The simulated environment is a replica of the University of New South Wales Canberra campus, shown in Fig. \ref{fig:budget_exp2}, where light blue objects represent the real static obstacles. A UGV team comprising 5 Jackal mobile UGVs is used.

% \begin{figure}
% 	\centering
% 	\begin{subfigure}[b]{0.35\textwidth}
% 		\includegraphics[width=\textwidth]{Figs/simulated_env_34}
% 		\label{fig:real_en}
% 	\end{subfigure}
% 	\caption{\footnotesize Experiments 2-4: Environment for characterising the path following performance.}\label{fig:urban_en}
% \end{figure}

Each experiment is repeated 5 times. Mean and 95$\%$ confidence intervals are reported where appropriate. The $P$-Value from the Student T-test is used to distinguish the performance of different algorithms. If the variance of means between two sets is less than the expected $P$ value of 0.05 (i.e., 5$\%$), we assume the results are statistically significant. The initial locations of the five vehicles are varied after every trial and given in Table \ref{tab:pose_jackal}:

\begin{table}[h]
 \centering
\caption {Experiments 2-4: Initial locations of UGVs} 
\label{tab:pose_jackal}
  \centering
  \resizebox{0.49\textwidth}{!}{%
  \begin{tabular}{llll}
    \hline
     \textbf{Trial Num} &  \textbf{UGV 1} &  \textbf{UGV 2}  &  \textbf{UGV 3}  \\
      \hline
1 & $[-23~0] \pm 0^o$ & $[-26~-1] \pm 0^o$ & $[-26~0] \pm 0^o$ \\
2 & $[-23.05 ~0.05] - 10^o$ & $[-26.05~-1.05] - 10^o$ & $[-26.05~0.05] - 10^o$  \\
3 & $[-22.95 ~-0.05] + 10^o$ & $[-25.95~-0.95] + 10^o$ & $[-25.95~-0.05] + 10^o$  \\
4 & $[-23.1 ~0.1] - 20^o$ & $[-26.1~-1.1] - 20^o$ & $[-26.1~1] - 20^o$   \\
5 & $[-22.9 ~-0.1] + 20^o$ & $[-25.9~-0.9] + 20^o$ & $[-25.9~-0.1] + 20^o$   \\
\hline
 \textbf{Trial Num} & \textbf{UGV 4}  &  \textbf{UGV 5} & \\
      \hline
1 & $[-26~1] \pm 0^o$ & $[-26~2] \pm 0^o$ & \\
2 & $[-26.05~1.05] - 10^o$ & $[-26.05~2.05] - 10^o$ & \\
3 & $[-25.95~0.95] + 10^o$ & $[-25.95~1.95] + 10^o$ & \\
4 & $[-26.1~1.1] - 20^o$ & $[-26.1~2.1] - 20^o$ & \\
5 & $[-25.9~0.9] + 20^o$ & $[-25.9~1.9] + 20^o$ & \\
\hline
  \end{tabular}}
\end{table}

Each UGV's maximum linear speed is 2.0$m/s$, and the maximum angular velocity is set to $\pm1.5rad/s$. Table \ref{tab:flocking} summarises all parameter settings for the simulation experiments.

\begin{table}[h]
 \centering
\resizebox{0.48\textwidth}{!}{%
\begin{tabular}{llll}
    \hline
     \textbf{Param} &  \textbf{Unit} & \textbf{Description} &   \textbf{Jackal}\\
      \hline
$\omega_{c}$ & - & Weight for cohesion rule & 0.27\\
$\omega_{a}$ & - & Weight for alignment rule  & 1.05\\
$\omega_{s}$ & - & Weight for separation rule & 1.65\\
$\omega_{w}$ & - & Wall Weight & 1.1  \\
$\omega_g$ & - &  Goal Weight  & 1.3 \\
$\omega_{cx}$ & - &  X-Axis Weight  & 0.35 \\
$\omega_{cy}$ & - &  Y-Axis Weight  & 0.35 \\
$\omega_{c\theta}$ & - &  Yaw-Angle Weight  & 0.3 \\ 
$R_{a}$ & (m) & Alignment Radius & 5 \\
$R_{s}$ & (m)&  Separation Radius & 1.1 \\
$R_{c}$ & (m)&  Cohesion Radius & 5 \\
$R_{av}$ & (m)&  Avoiding Radius & 0.3 \\
$R_d$ & (m)&  Obstacle detection range & 1.6 \\
$k_0$ &  Spring coefficient & 5.0 \\
$\delta_d$ &  Distance tolerance & 0.05m  \\
$\delta_\theta$ &  Angle tolerance & 15\degree  \\
$\omega_{g}$ &  Target force weight & 1.1 \\
$v_{q_0}$ &  Virtual leader's speed & 1.1 \\
\hline
  \end{tabular}}
  \caption {Experiments 2-4: Parameters of flexible formation control and the comparative frontier-led swarming algorithm.} 
\label{tab:flocking}
\end{table}

The maximum coverage path length and coverage time  ($L_{max},T_{max}$) are 3500m and 30000s. \\

\subsubsection{Experiments 2-4 Performance Metrics}
In this series of experiments, the following metrics are examined:
\begin{itemize}
    \item Coverage percentage ($CP$), the percentage of the environment visited by the UGVs
    \item Path length ($PL$), the length of the coverage path
    \item Turnaround time ($TT$), the time to follow the coverage path (or achieve 100\% coverage in the case of the FS algorithm)
    \item Coverage redundancy ($CR$). Coverage redundancy (backtracking over areas already covered) can be calculated using (\ref{eq:cr_metric}) where $C_{repeated}$ is the average number of all repeated coverage cells.
    \item Group ($G$), how close together the UGVs are, defined as (\ref{eq:group_metric}) where $N_{s}$ is the number of swarming UGVs, $N_t = 5$ is the number of trials. $\bar{p}_{a}$ is the average position of the involved UGVs at the given time. We evaluate the group and order every 500 steps, as an average over the proceeding 150-time steps.
    \item Order ($O$), how well-aligned UGVs are in terms of both speed and direction as defined in \ref{eq:order_metric}) where $\bar{\varepsilon}_{a}$ is the average velocity of the involved UGVs at the given time. We also evaluate ordering every 500 steps as an average over the proceeding 150-time steps.
\end{itemize}

\begin{equation}\label{eq:cr_metric}
CR = \frac{|C_{repeated}|}{|C_{free}|} \times 100\%,
\end{equation}

\begin{equation}\label{eq:group_metric}
G = \frac{\sum_{i=0}^{N_t} \sum_{t=T_0}^{TT} \frac{\sum_{i=1}^{N_{s}} ||p^{i} - \bar{p}_a||_2}{TT-t}}{N_{s}~N_t},
\end{equation}

\begin{equation}\label{eq:order_metric}
O = \frac{\sum_{i=0}^{N_t} \sum_{t=T_0}^{TT} \frac{\sum_{i=1}^{N_{s}} ||\varepsilon^{i} - \bar{\varepsilon}_a||_2}{TT-t}}{N_{s}~N_t},
\end{equation}

\subsubsection{Experiment 2 Group and Order results} 
\begin{figure}
	\centering
	\begin{subfigure}[b]{0.5\textwidth}
		\includegraphics[width=\textwidth]{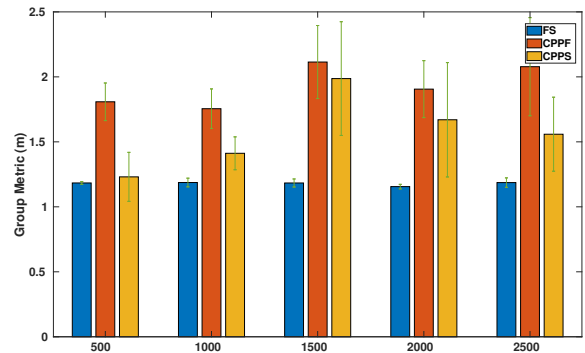}
		\caption{\footnotesize \textit{Group Metric.}}
		\label{fig:CP_hete_scenario1}
	\end{subfigure}
	\begin{subfigure}[b]{0.5\textwidth}
		\includegraphics[width=\textwidth]{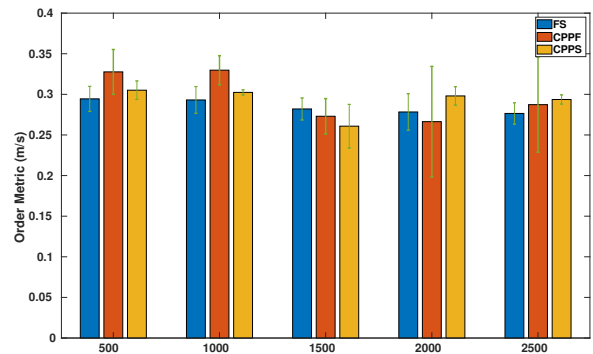}
		\caption{\footnotesize \textit{Order Metric.}}
		\label{fig:ACP_hete_scenario1}
	\end{subfigure}
	\caption{\footnotesize Experiment 2: Group and order metrics attained by each algorithm at different times during the coverage task.}\label{fig:swarming_exp0}
\end{figure}

First, we examine the ability of all three algorithms to keep the UGVs together and heading in the same direction and at the same speed. Fig. \ref{fig:swarming_exp0} indicates that all the approaches maintain a reasonable level of grouping and ordering relative to the search space size. The FS approach maintains a tighter formation than the CPPF and CPPS approaches in all experimental settings. The difference in group metrics is statistically significant at the 95\% confidence level. This is likely due to the influence of the block-size switches. However, grouping and order are still maintained by the CPPF and CPPS methods, and the group and order metrics do not change significantly over time. The order metrics are similar for all three approaches. This makes sense because the swarming and formation approaches should both encourage ordering.

%Three approaches can obtain the group metrics with minimal deviations (approximately $\pm0.15m$) from the desired separation distances for the queuing formation (1.8m), inverted V and U formations (2.1m), and swarm (1.1m). The order metrics exhibit minor errors compared to 0. Moreover, the two metrics do not change significantly over time. This makes sense because the swarming and formation approaches should both encourage ordering.

\subsubsection{Experiment 3 Coverage performances with three different speed caps} 
Three maximum vehicle speeds are set as 0.45m/s (low), 0.7m/s (medium), 0.95m/s (high) in this experiment. This permits us to examine whether high speeds cause the UGVs to miss covering parts of the environment. Based on the default coverage path length budget, turnaround time budget, and the maximum linear and rotation speeds, the optimisation algorithm proposes a grid cell size of 2.25m that should result in a coverage path with $\hat{L}$ of 2007.34m and $\hat{T}$ of 5039.04s for the low speed; 3445.92s for the medium speed and 2991.28s for the high speed. $\hat{CP}$ should be 100\%. The LiDAR sensor's observation range is 4m. 

\begin{figure}
	\centering
	\begin{subfigure}[b]{0.46\textwidth}
		\includegraphics[width=\textwidth]{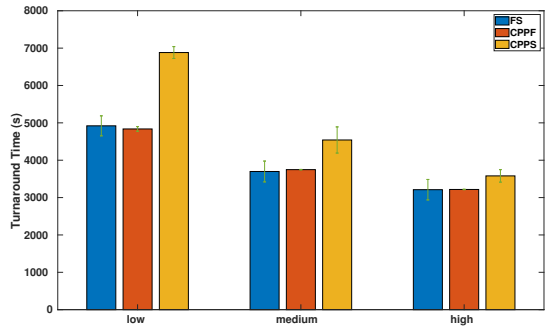}
		\caption{\footnotesize \textit{Turnaround Time.}}
		\label{fig:tt_exp2}
	\end{subfigure}
	\begin{subfigure}[b]{0.5\textwidth}
		\includegraphics[width=\textwidth]{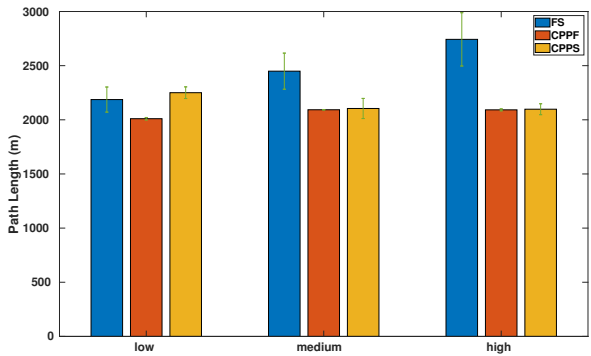}
		\caption{\footnotesize \textit{Path Length.}}
		\label{fig:pl_exp2}
	\end{subfigure}
	\begin{subfigure}[b]{0.5\textwidth}
		\includegraphics[width=\textwidth]{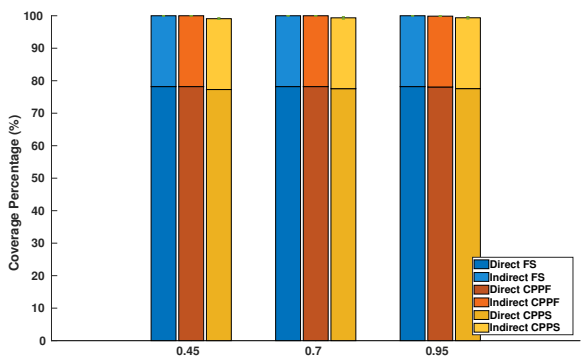}
		\caption{\footnotesize \textit{Coverage Percentage.}}
		\label{fig:cp_exp2}
	\end{subfigure}
		\begin{subfigure}[b]{0.5\textwidth}
		\includegraphics[width=\textwidth]{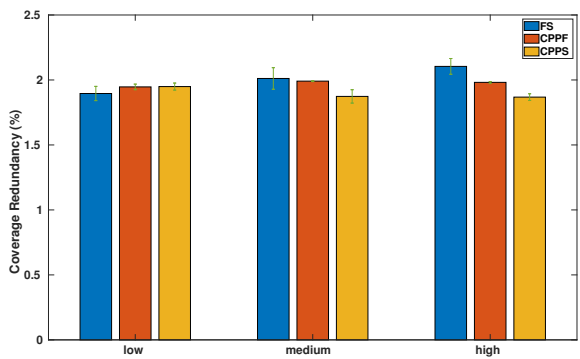}
		\caption{\footnotesize \textit{Coverage Redundancy.}}
		\label{fig:cr_exp2}
	\end{subfigure}
	\caption{\footnotesize Experiment 3: Turnaround time, path length, coverage percentage, and coverage redundancy metrics for three different speed caps.}\label{fig:metric_exp2}
\end{figure}

Fig. \ref{fig:metric_exp2} shows the turnaround time, path length, coverage percentage, and coverage redundancy metrics. 

As expected, the higher the maximum vehicle speed, the shorter the turnaround time is. Using CPPF and FS, the TTD is below 300s, while TTD for CPPS is approximately double. CPPF and FS strategies were also able to cover the whole region ($78.19\pm0.00\%$ direct $CP$ plus $21.81\pm0.00\%$ indirect $CP$), whereas the CPPS achieved an incomplete coverage percent ($77.37\pm0.32\%$ for the direct coverage and $21.99\pm0.00\%$ for the indirect coverage). This is because CPPS uses an organic formation control strategy that is unable to backtrack. 

CPPF has a PLD of under 90m, while FS has the largest PLD. This is because FS (which does not include a planning component) often needs to backtrack to cover a missed area. The differences in CR for the comparative algorithms are insignificant at the 95\% confidence level, except in the high-speed case, where the FS produces the highest redundant coverage percentage, namely, $2.1\pm0.06\%$. While the path planning-based coverage methods generate the non-backtracking paths, the frontier search technique guides the physical UGVs to track frontier cells, leading to an increase in back-tracking the covered areas.

% The main reason can be seen in Fig. \ref{fig:multi_UGV_paths}. While the path planning-based coverage methods generate the non-backtracking paths, the frontier search technique guides the physical UGVs to track frontier cells, leading to an increase in back-tracking the covered areas.

% \begin{figure}
% 	\centering
% 	\begin{subfigure}[b]{0.5\textwidth}
% 		\includegraphics[width=\textwidth]{Figs/paths_FS.eps}
% 		\caption{\textit{Frontier-led swarming}}
% 		\label{fig:paths_FS}
% 	\end{subfigure}
% 	\begin{subfigure}[b]{0.5\textwidth}
% 		\includegraphics[width=\textwidth]{Figs/paths_CPPS.eps}
% 		\caption{\textit{Coverage path planning with swarm following.}}
% 		\label{fig:paths_CPPS}
% 	\end{subfigure}
% 	\begin{subfigure}[b]{0.5\textwidth}
% 		\includegraphics[width=\textwidth]{Figs/paths_CPPF.eps}
% 		\caption{\textit{Coverage path planning with flexible formation control.}}
% 		\label{fig:paths_CPPF}
% 	\end{subfigure}
% 	\caption{\footnotesize Experiment 3: UGV paths resulting from the use of different algorithms in a high budget scenario.}\label{fig:multi_UGV_paths}
% \end{figure}

\subsubsection{Experiment 4 Performance with different time and path length budgets}
All algorithms were also run with three different time and path length budgets: (1) high budgets of $[4000s, 2500m]$ (more than enough time for all algorithms to achieve 100\% coverage), (2) tight budgets of $[2800s, 1800m]$ (barely enough time for one of the algorithms to finish), (3) very tight budgets of $[1200s, 800m]$ (no algorithm will be able to achieve optimal solution-approximately 77.15\% map covered). 

\begin{figure}
	\centering
	\begin{subfigure}[b]{0.43\textwidth}
		\includegraphics[width=\textwidth]{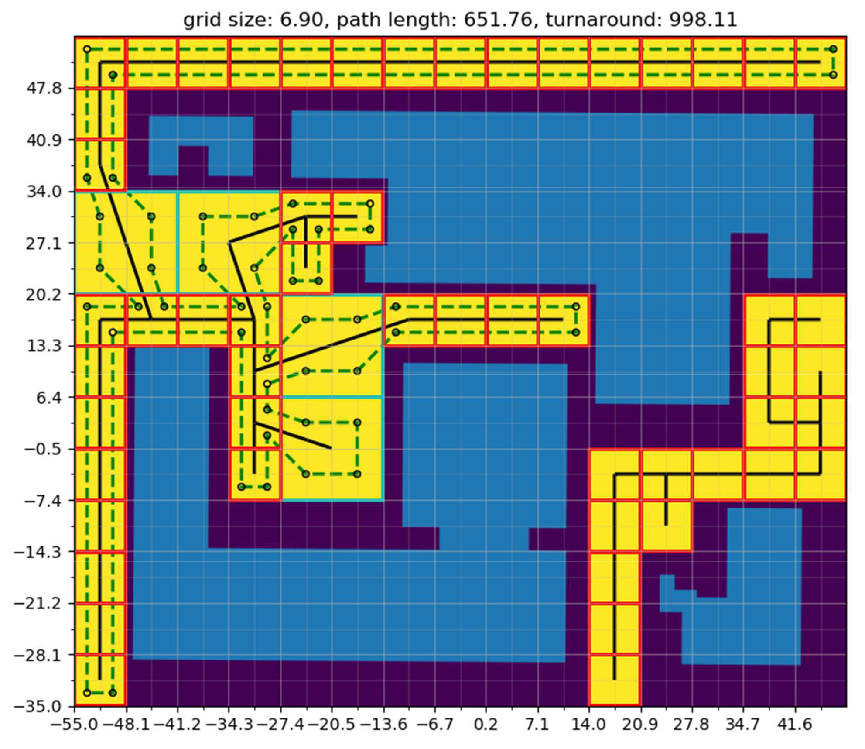}
		\caption{\footnotesize \textit{Very Tight Budget.}}
		\label{fig:tt_exp2}
	\end{subfigure}
	\begin{subfigure}[b]{0.45\textwidth}
		\includegraphics[width=\textwidth]{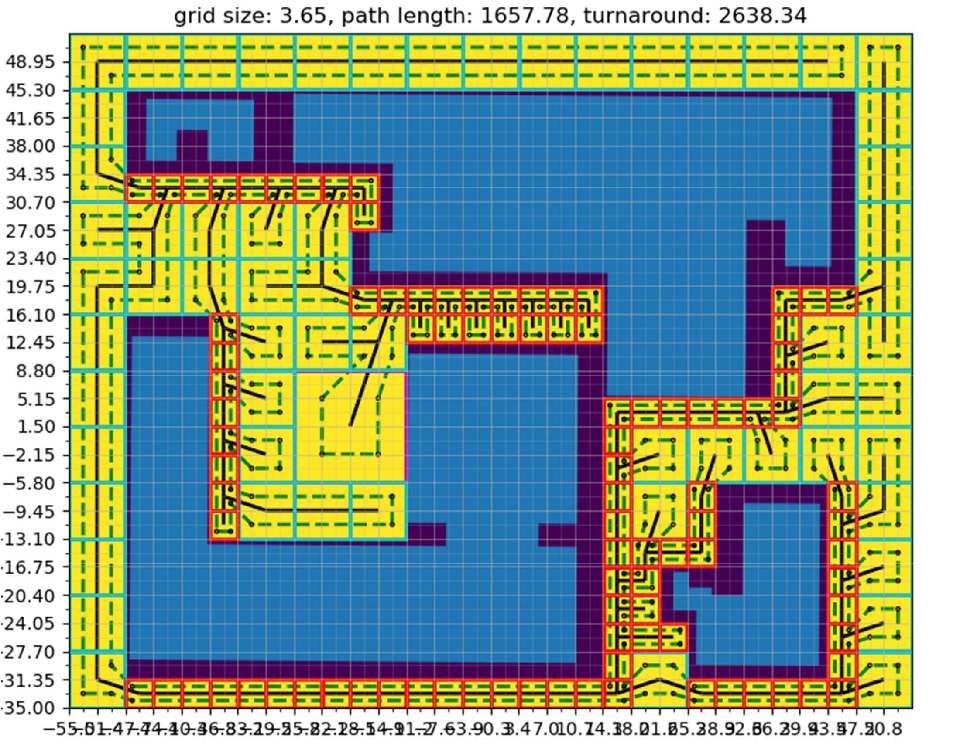}
		\caption{\footnotesize \textit{Tight Budget.}}
		\label{fig:pl_exp2}
	\end{subfigure}
	\begin{subfigure}[b]{0.45\textwidth}
		\includegraphics[width=\textwidth]{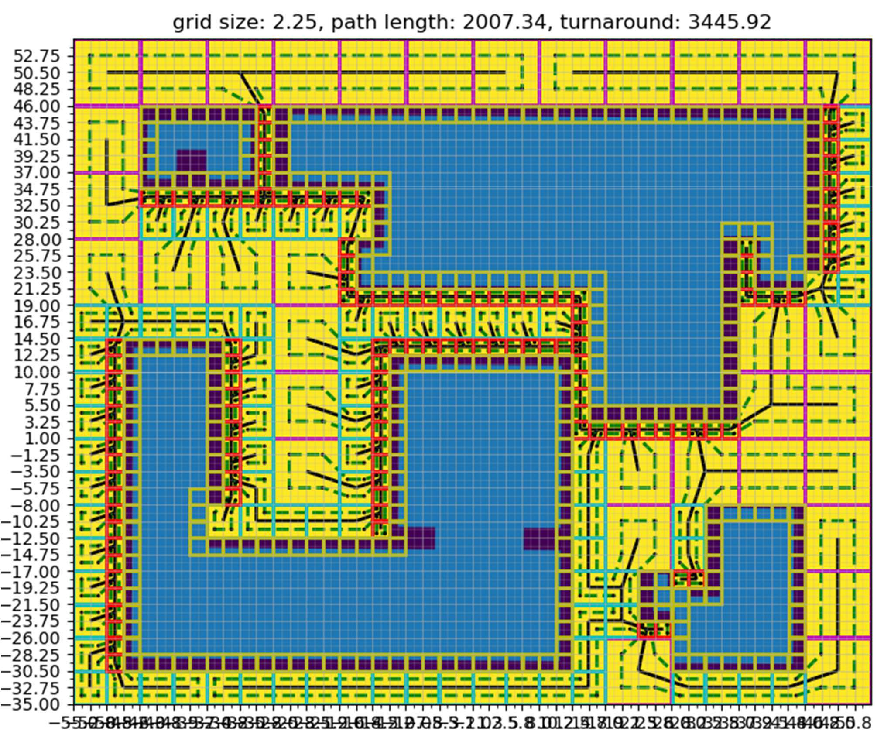}
		\caption{\footnotesize \textit{High Budget.}}
		\label{fig:cp_exp2}
	\end{subfigure}
	\caption{\footnotesize Experiment 4: Coverage path predictions made for different budgets.}\label{fig:budget_exp2}
\end{figure}

The planner produces the paths shown in Fig. \ref{fig:budget_exp2} for each of these cases. In these figures, blue polygons show the real obstacle positions. Purple cells indicate regions denoted as obstacles as a result of different grid cell size recommendations. Yellow cells represent open areas. All UGVs' maximum speed in all experiments is set at 0.7m/s.
   
\begin{figure}
	\centering
	\begin{subfigure}[b]{0.45\textwidth}
		\includegraphics[width=\textwidth]{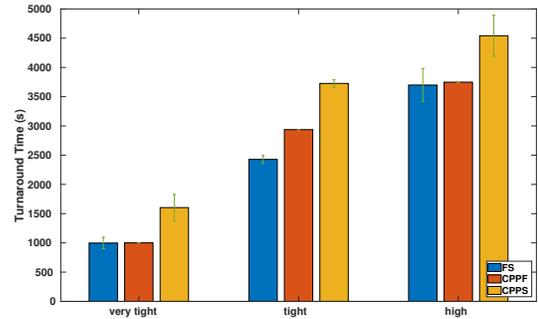}
		\caption{\footnotesize \textit{Turnaround Time.}}
		\label{fig:tt_exp3}
	\end{subfigure}
	\begin{subfigure}[b]{0.45\textwidth}
		\includegraphics[width=\textwidth]{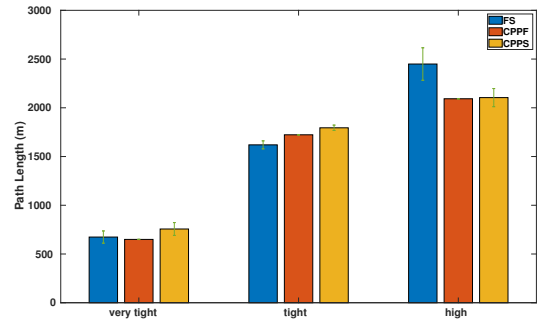}
		\caption{\footnotesize \textit{Path Length.}}
		\label{fig:pl_exp3}
	\end{subfigure}
	\begin{subfigure}[b]{0.45\textwidth}
		\includegraphics[width=\textwidth]{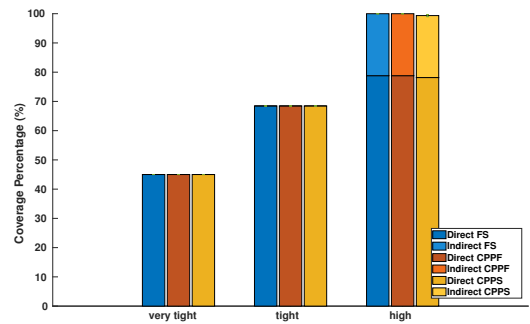}
		\caption{\footnotesize \textit{Coverage Percentage.}}
		\label{fig:cp_exp3}
	\end{subfigure}
		\begin{subfigure}[b]{0.45\textwidth}
		\includegraphics[width=\textwidth]{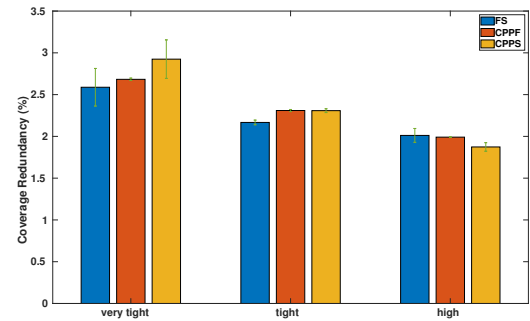}
		\caption{\footnotesize \textit{Coverage Redundancy.}}
		\label{fig:cr_exp3}
	\end{subfigure}
	\caption{\footnotesize Experiment 4: Turnaround time, path length, coverage percentage, and coverage redundancy metrics for three different budgets.}\label{fig:metric_exp3}
\end{figure}

In general, Fig. \ref{fig:metric_exp3} demonstrates that the CPPF strategy presents minimal offsets from the desired coverage parameters (such as the maximum coverage percentage, total path length, and coverage time) in both cases. The FS and CPPF exhibit quite similar metrics for the very tight and high budgets scenarios. In contrast, the FS has a superior $TT$ figure for the tight-budget case, with the lowest $PL$, and $CR$ values. This difference is statistically significant at the 95\% confidence level. Moreover, there are no significant differences in the $TT$ and $CR$ parameters when high budgets are allowed.

As shown in Fig. \ref{fig:cp_exp3}, it is interesting to see that both strategies only achieve partial (direct) coverage in the tight and very tight-budget scenarios because all obstacle boundaries are not observed by the LiDAR sensor (its observation range is smaller than the grid cell size). They produce the same direct $CPs$ of only $45\pm0.00\%$ and $68.46\pm0.00\%$. However, they generate full coverage, including direct and indirect coverage, when a high budget is employed. The $CP$ metric of the CFFS is $0.2\%$ lower than those of the CPPF and FS. The reason is that the higher the desired budget is set, the smaller the grid cell size is. In the high-budget case, the grid cell size of 2.25m is significantly less than LiDAR's observation range of 4m.

\subsection{Experiment 5: Comparison with a Quadtree Data Structure}

\begin{figure}
	\begin{center}
		\begin{tabular}{cc}	
			\includegraphics[width=15.5pc]{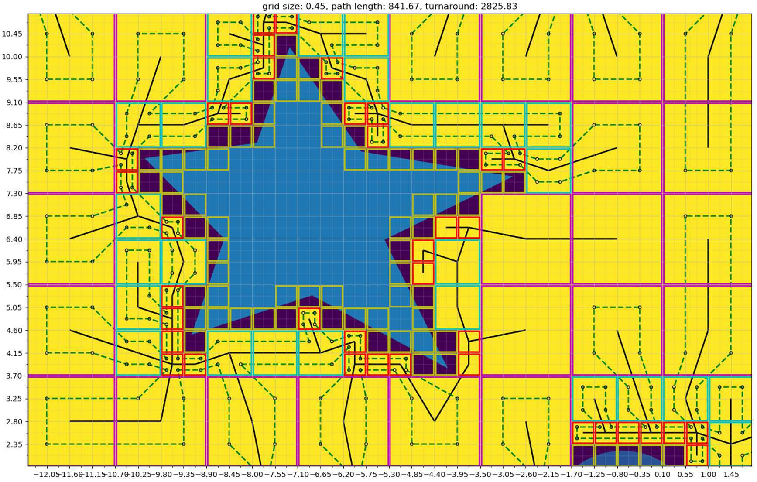} \\
			(a) \textit{Quadtree spanning tree.}\\[6pt]
			\includegraphics[width=15.5pc]{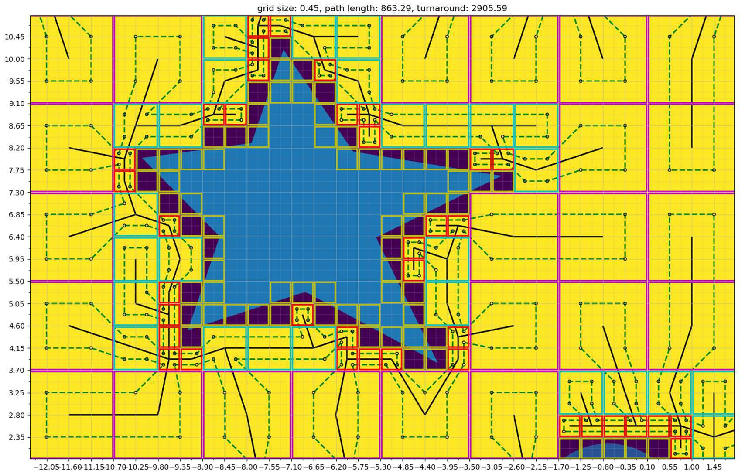} \\
			(b) \textit{CPPF spanning tree.} 
		\end{tabular}
		\caption{\footnotesize  Spanning trees produced by CPPF and \ Quadtree approaches. Compared to the CPPF method, the Quadtree method cannot build the robot path (green-dotted line) crossing the eleven yellow cells (located on the right of the star obstacle). See Fig. \ref{fig:qd_connection} for details.}
		\label{fig:map_exp4}
	\end{center}
\end{figure}

The spanning tree produced by the multi-resolution block structure used in CPPF can have a significant impact on coverage results. To assess this impact, we compared the performance of MCPP in CPPF with that of a conventional homogeneous quadtree algorithm, where all blocks are of uniform size (0.45m), using a 25m $\times$ 25m grid map containing oddly shaped obstacles as illustrated in Fig. \ref{fig:map_exp2}. Complex-shaped obstacles can cause a block to have multiple neighbor blocks in the same direction, resulting in incomplete coverage routes (green line) and discontinuous minimum spanning trees (black line) in the quadtree method. However, our MCPP algorithm resolves this issue in the third case of Algorithm \ref{alg:pp_algo} (Lines 12-24). In this case, when obstacle cells exist between two adjacent blocks, the joint point still satisfies the condition of being within the intersection region between the considered block and the triangle formed by the three centers of the considered block and the two adjacent blocks (see Fig. \ref{fig:map_exp4}).

\begin{figure}[H]
    \centering
    \includegraphics[scale=0.43]{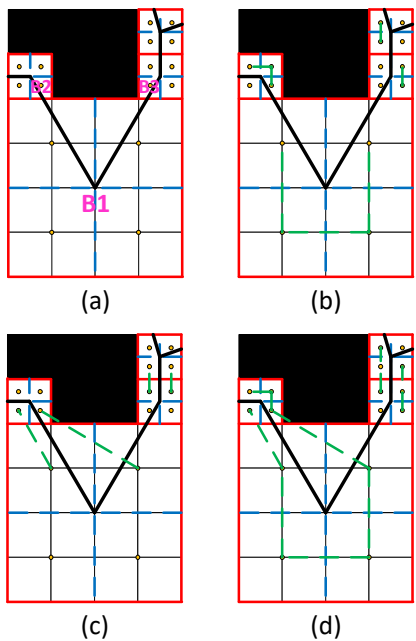}
    \caption{\footnotesize The quadtree procedure for connecting the robot path around MST. (a) The MST; (b) The coverage path inside cells; (c) The coverage path between cells; (d) The incomplete coverage path.}
    \label{fig:qd_connection}
\end{figure}

\begin{figure}[H]
    \centering
    \includegraphics[scale=0.43]{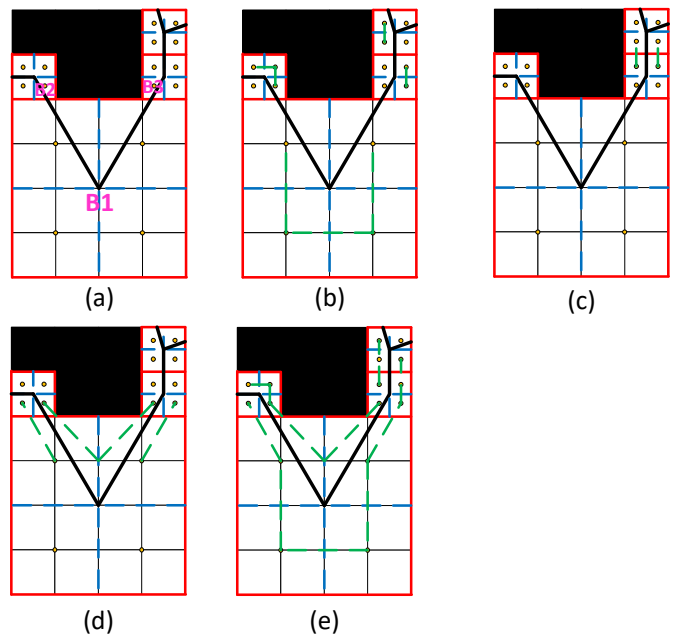}
    \caption{\footnotesize The CPPF procedure for connecting the robot path around MST. (a) The MST; (b) The coverage path inside cells; (c) The coverage path between cells; (d-e) The complete coverage path.}
    \label{fig:mcpp_connection}
\end{figure}

As observed in Figs. \ref{fig:qd_connection} and \ref{fig:mcpp_connection}, blocks B2 and B3 are located on top of block B1, as determined by the MST. The quadtree algorithm is then applied to link the two upper parts of B1 with the two lower parts of B2, thereby precluding the possibility of establishing an additional connection with B3. The disruption of the connection between B1 and B3 would result in the loss of the remaining branch of the MST. In the absence of obstacles on the map, the minimum spanning tree will exhibit, at most, a single connection between any two blocks in any given direction, and the seamless continuity of the path is ensured by the quadtree algorithm. Our proposed algorithm establishes a path that connects B1 to both B2 and B3, resulting in a path that guarantees that all parts of the MST are followed to form the robot path.

Furthermore, although all estimates, such as maximum coverage percentage, total path length, and coverage time, can be well achieved by both methods when combined with our formation control solution, as demonstrated in Experiments 1-4, the coverage percentage obtained by running the quadtree algorithm is lower than our spanning tree method with modified rules. This is because the multi-robot team cannot move through the blocks, such as the eleven yellow blocks shown in Fig. \ref{fig:map_exp4}, to update the coverage information due to the absence of the coverage paths.

% we compare the performance of path planning to a traditional homogeneous quadtree algorithm in which all blocks are of uniform size (0.45m). We make this comparison on the 25m$\times$25m grid map shown in Fig. \ref{fig:map_exp2}. This environment is characterised by a number of oddly shaped obstacles. When obstacles are present, such as at the bottom vertex of the star obstacle, a block may have multiple neighbor blocks in the same direction, resulting in discontinuous MSTs (black line) and incomplete coverage routes (green line) in the QD method. \red{However, the MCPP algorithm provides a solution for this issue in its third case (Line 12-24 of Algorithm \ref{alg:pp_algo}). In this case, where obstacle cells exist between two adjacent blocks, the joint point still satisfies the condition of being within the intersection region between the considered block and the triangle formed by the three centers of the considered block and the two adjacent blocks.}

% Fig. \ref{fig:map_exp4} indicates that although all estimates, such as the maximum coverage percentage, total path length, and coverage time, can be still performed well by both methods, the MST, the coverage path, and the coverage percentage obtained by running the QD are significantly worse than those of the CPPF method as they cannot bypass three yellow cells to update the coverage information. 
%Figure \ref{fig:map_exp4} indicates that

\subsection{Experiment 6: Outdoor experiments using Jackals and DGPS positioning}

\subsubsection{Experiment 6 Outdoor Setup}

\begin{figure}
	\centering
	\begin{subfigure}[b]{0.49\textwidth}
		\includegraphics[width=\textwidth]{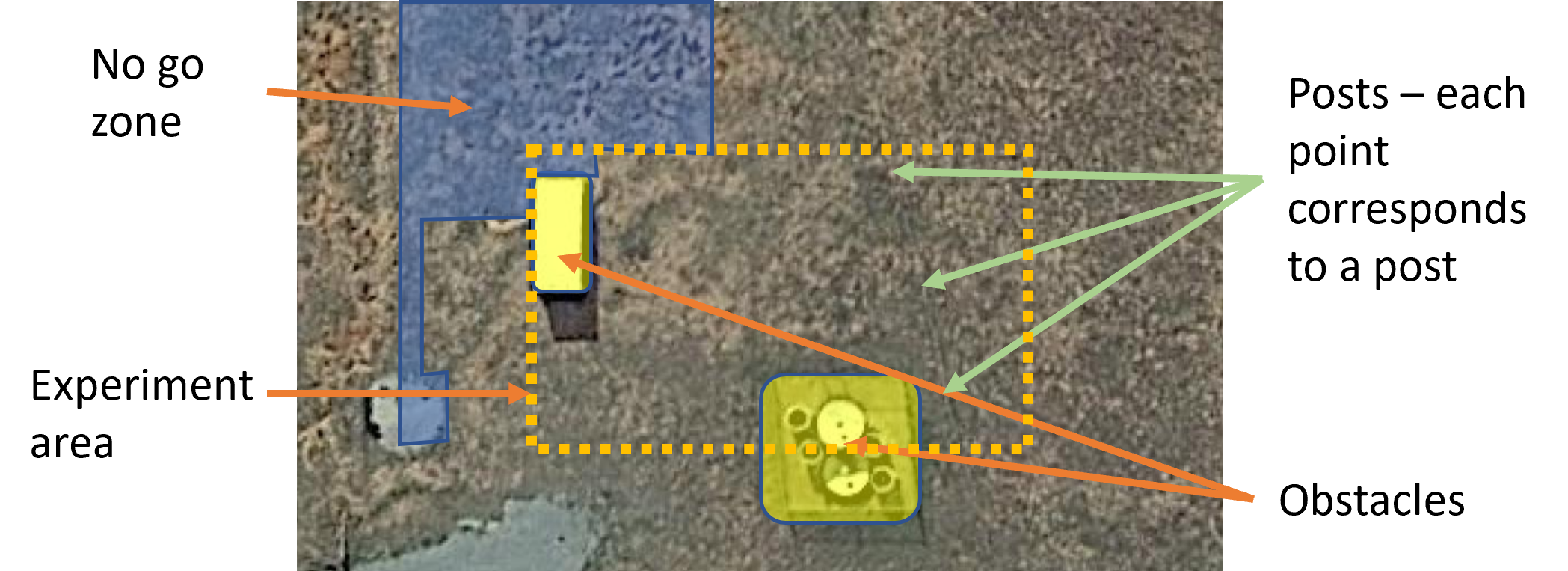}
	\end{subfigure}
	\caption{\footnotesize Experiment 6: Aerial view of the field site.}\label{fig:real_world}
\end{figure}

To verify the effectiveness of the algorithms on real UGVs, a $15.25m \times 24.4m$ outdoor setup was used involving several arbitrary-shaped obstacles (see Fig. \ref{fig:real_world}) and a team of three Jackal UGVs.  There were three significant obstacles: a hut, a shipping container, and a piece of equipment covered with a metal mesh cage. Several steel posts were on the terrain, each with a Y cross-section. The diameter of a post (assuming a circular cross-section) was roughly 6cm. In addition to these fixed obstacles, a no-go zone for the UGVs was included to prevent the UGVs from crossing over known rabbit burrows. 

The Jackal UGVs used in our experiments are fitted with a differential GPS (DGPS) rover module, Inertial Measurement Unit (IMU) and a SICK LMS-111 LiDAR. The LiDAR's observation range is 4m. Incoming GPS rover signals and the internal IMU sensor were read at a rate of 10Hz. The DGPS base station was stationary and transmits DGPS corrections to the rovers. The sampling time of the whole system was set at 30Hz. The role of the ROS Master is to enable individual ROS nodes to locate one another. Unlike the traditional implementations that run the ROS Master on only a single base station, the ROS Master was implemented on each UGV to facilitate ROS communications and improve the system's robustness.

Five trials were conducted. The problem constraints given were the maximum path length $L_{max}$ of 100m, and maximum coverage time $T_{max}$ of 600s. For each test, the UGVs were initially setup in a V-formation close to the bottom-left corner at GPS-measured coordinates of [9.53, 15.58, 0]m, with the formation facing down towards the bottom of the map. Some slight variations in initial UGV locations were introduced at the start of each test. The initial locations of the three vehicles are given in the columns of (\ref{eq:ini_pos1}). The virtual leader's forward speed $v$ is initialised to 0.2$m/s$. The maximum turn rate when maneuvering is set to $0.73rad/s$. Other parameters of our experiments are summarised in Table \ref{tab:formation_control}.
\begin{equation}\label{eq:ini_pos1}
q_{p_i}(0) = \begin{bmatrix}
9.23 & 10.01 & 8.07 \\
15.69 & 18.4 & 18.32  \\
0   &  0 & 0   
\end{bmatrix}
\end{equation}

\begin{table}[h]
 \centering
\caption {Experiment 6: Parameters of leader-follower formation control} 
\label{tab:formation_control}
  \centering
  \begin{tabular}{lll}
    \hline
     \textbf{Parameter} &  \textbf{Description} &  \textbf{Jackal} \\
      \hline
$k_0$ &  Spring coefficient & 3.9 \\
$\zeta$ &  Vision range of LiDAR sensor & 4.0m \\
$\delta_d$ &  Distance tolerance & 0.3m  \\
$\delta_\theta$ &  Angle tolerance & 15\degree  \\
$R_{av}$ &  Avoidance radius & 0.32m  \\
$\omega_{g}$ &  Target force weight & 1.1 \\
\hline
  \end{tabular}
\end{table}

\subsubsection{Experiment 6 Performance metrics}
The metrics used in this experiment were: (1) difference between the actual and predicted path length (PLD); (2) difference between the actual and predicted turnaround time (TTD); (3) Coverage percentage (CP); (4) coverage redundancy (CR); (5) group (G); and (6) order (O).
\\
\subsubsection{Experiment 6 Results}
\begin{figure}
	\begin{center}
		\begin{tabular}{cc}	
			\includegraphics[width=15.5pc]{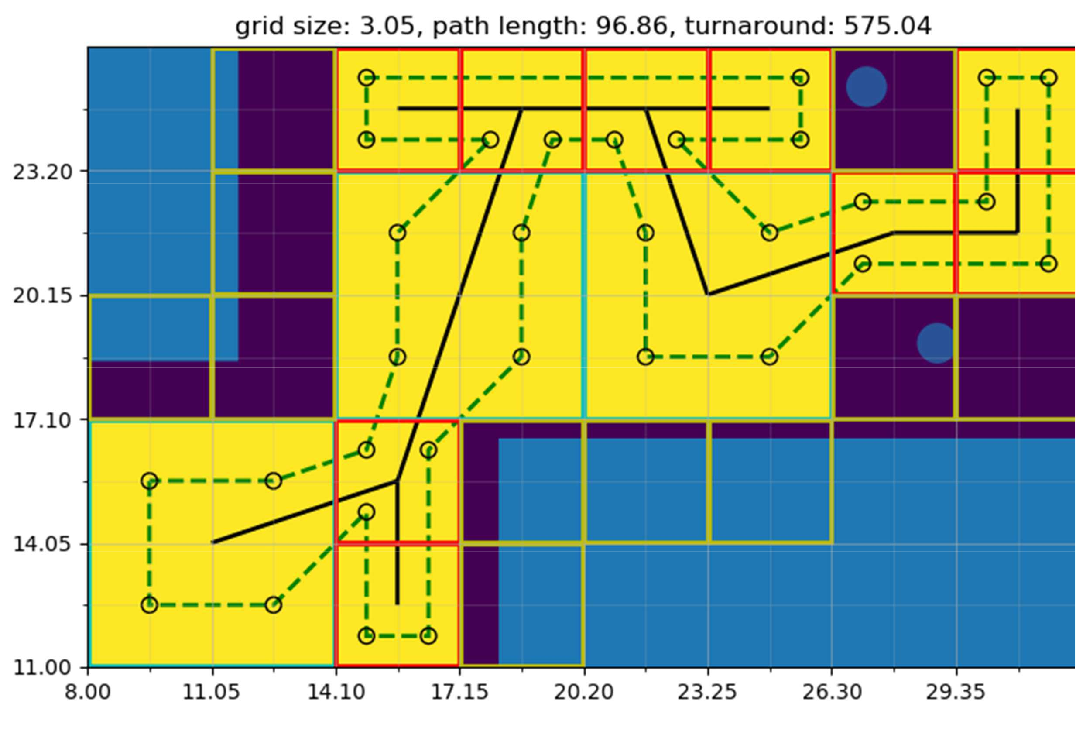} \\
		\end{tabular}
		\caption{\footnotesize Experiment 6: Predicted path for the virtual leader in the outdoor field.}
		\label{fig:real_pred_vicon}
	\end{center}
\end{figure}

The experimental test performed can be viewed in the following videos: \url{https://youtu.be/z4kK6OnXXg8}. The prediction engine took 2s to produce the recommended path shown in Fig. \ref{fig:real_pred_vicon}. It recommends a grid cell size of $3.05m$, and predicts the path length $\hat{L}$ of 96.86m, coverage time $\hat{T}$ of 575.04s, and coverage percentage $CP$ of 100$\%$. 

As shown in the video, the map is entirely covered (namely $90\pm0.00\%$ for the direct coverage and $10\pm0.00\%$ for the indirect coverage) after $586.17\pm1.2$s (mean over five tests $\pm$ standard deviation). The travelled path length is approximately $103.03\pm0.9$m. These results reflect a PLD of 6.17m and TTD of 11.13s. These results are slightly higher than those from our simulations. We observed this is due to the variations in the initial UGV position among five tests and the actual tracking errors caused by the terrain's uneven round with thick grass and the DGPS and IMU sensor errors. On the other hand, the advantage of using the spanning tree to design the complete coverage paths can be seen through the low CR of $7.33\pm1.41$ when the back-tracking routes are eliminated, and only several cells at the corners are covered repeatedly. 

\begin{figure}
\centering
	\includegraphics[width=23.5pc]{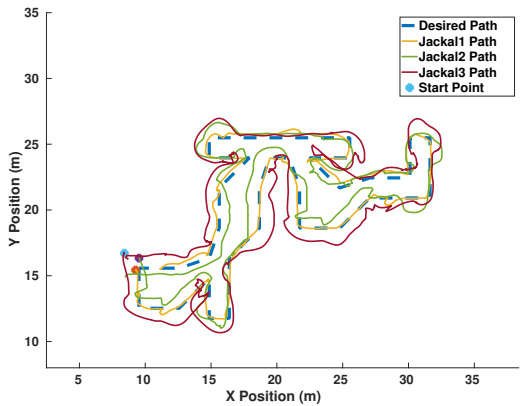}
	\caption{\footnotesize Experiment 6: Planned path versus actual paths produced by three UGVs in one run.}
	\label{fig:real_path}
\end{figure}

As shown in Fig. \ref{fig:real_path}, there are slight fluctuations in the actual paths. This is due to three factors. First, we measured GPS static position errors of 0.08m that contribute to some deviations in movement. Secondly, as the robots switch between V, U and queuing formations, there is some distortion in the paths. Finally, inter-UGV collision avoidance exerts influence on the robots causing deviations from the path.

\begin{figure}
	\centering
	\begin{subfigure}[b]{0.48\textwidth}
		\includegraphics[width=\textwidth]{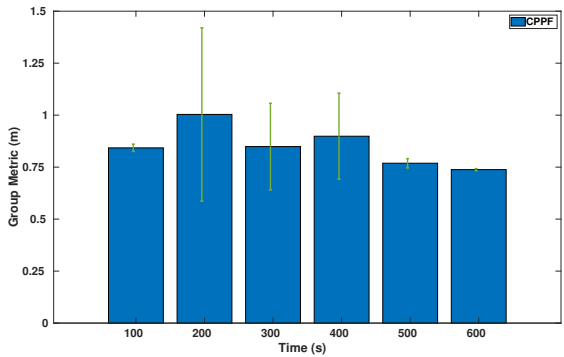}
		\caption{\footnotesize \textit{Group.}}
		\label{fig:group_exp4}
	\end{subfigure}
	\begin{subfigure}[b]{0.48\textwidth}
		\includegraphics[width=\textwidth]{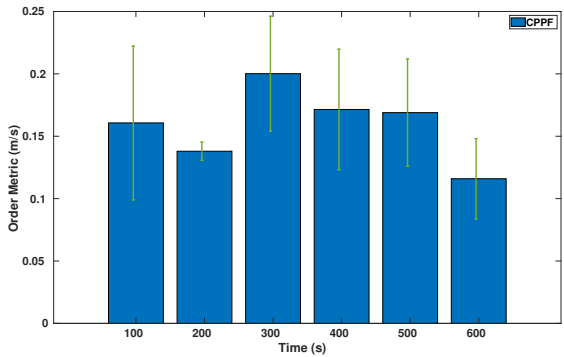}
		\caption{\footnotesize \textit{Order.}}
		\label{fig:order_exp4}
	\end{subfigure}
	\caption{\footnotesize Experiment 6: Formation metrics for the real-world experiments.}\label{fig:swarm_metric_exp4}
\end{figure}

However, under the control of our formation and role assignment strategies, three Jackal follower UGVs switch positions and effectively form the desired formations when tracking the virtual leader's motion and avoiding obstacles along the designed coverage path. The UGV behaviors exposed in the real-time environment are similar to those obtained in simulations. As a result, the proposed methods yield reasonable $G$ and $O$ figures, although there are short periods where there is high variance in the grouping. This occurs when one UGV strays away from the others (e.g. as a result of a brief increase in GPS error). The system self-corrects when GPS is re-acquired. Additionally, there are no significant variations of these two variables over time (see Fig. \ref{fig:swarm_metric_exp4}).

%While the first metric indicates slight differences with respect to the desired separation distances of the queuing formation (0.85m) and the inverted V formation (1.1m), the second metric is nearly 0, with a maximum error of only $11.47\pm2.51$ degree/s. Additionally, there are no significant variations of these two variables over time (see Fig. \ref{fig:swarm_metric_exp4}). 

Besides stationary obstacles, dynamic obstacles such as humans can move unpredictably and obstruct a flock's movement along a planned path. However, by combining the MCPP method with the closest safe angle-based obstacle avoidance, our robot team was able to successfully navigate through a challenging environment without any communication. In a video demonstration (available at \url{https://youtu.be/UjihyOj-VCU}), the relevant UGVs quickly adjusted their path when the human obstacle violated their obstacle avoidance radius, safely steering towards the nearest safe area and then resuming their intended path. Furthermore, during formation switches, the UGVs were able to avoid mutual collisions. This work addresses the limitations of existing coverage path planning methods by incorporating obstacle avoidance techniques, which enable robots to navigate safely in a dynamic environment \cite{jang2020multi}.

\subsection{Summary of Experiments} 

To validate the effectiveness and efficiency of our algorithm for multi-agent systems, we conducted a comprehensive set of experiments in both simulated and real-world environments, with five trials for each experiment. The obtained results showed that our algorithm yielded the best coverage accuracy rate without collisions and a coverage redundancy rate of only 2.7\% when all physical limits were met. These results demonstrate our algorithm's high accuracy and efficiency in ideal conditions.

We encountered some limitations in the real-time field tests due to hardware and environmental factors and mobile obstacles. However, even under these challenging conditions, we still achieved similar results to those obtained in the simulated experiments. This indicates that our algorithm is robust and can perform and adapt well to changing real-world scenarios.

The significance of these results lies in the potential applications of multi-agent systems, such as search and rescue operations or environmental monitoring. Our algorithm provides an efficient and reliable method for coordinating multiple agents to explore a dynamically changing area with minimal overlap and using minimal resources.

\section{Conclusion}
This paper has described a novel approach to MCPP in a dynamic environment. The CPPF algorithm produces a coverage path that maximises the area of the environment that will be visited while meeting given time and path length budgets. Further, it permits the detection of unknown obstacles simultaneously with the steering of the mobile robot to avoid collisions. We demonstrated this algorithm on simulated and real Jackal UGVs. We provided a range of statistics showing the performance of the prediction engine and the path-following algorithms. We saw that:
We saw that:
\begin{itemize}
    \item The planner is able to recommend suitable grid cell sizes to meet given time and path length budgets within approximately 10s.
    \item Path length predictions are reasonably accurate for both simulated and real UGVs. The longer the necessary path, the lower the path length prediction error, with as little as 0.4\% error on a 1km path. 
    \item Turnaround time predictions range in accuracy from a few seconds over 100 metres to up to 5 minutes discrepancy on a 1km path.
    \item Formation-based path following achieves comparable coverage performance to a state-of-the-art reactive swarming approach but offers the guarantee of a time and path length prediction.
    \item It is feasible to use the algorithm on real UGVs in an outdoor setting.
\end{itemize}

The possibilities for future work in this area are rich. The proposed method constructs the optimal coverage path for every UGV using the MST such that the union of all paths generates a full coverage of the terrain. However, these paths are static, and the coverage is performed in static environments where the obstacles do not move. In future work, we will use a rapidly-exploring random tree (RRT) algorithm for local path re-planning to update the current path to achieve mobile obstacle avoidance in dynamic environments.

\section{Acknowledgement}
This work was supported by the Australian Defence Science and Technology Group (DSTG) under grant No. 9729.
\bibliographystyle{unsrt}
\bibliography{path_planning}

\end{document}